\documentclass[a4paper,fleqn]{cas-dc}

\usepackage[authoryear,longnamesfirst]{natbib}
\usepackage{algorithm}
\usepackage{multirow}
\usepackage{multicol}
\usepackage{makecell}
\usepackage{graphicx}
\usepackage{textcomp}
\usepackage{xcolor}
\usepackage{tabularx}
\usepackage{algpseudocode}
\usepackage{booktabs}
\usepackage{threeparttable}
\usepackage{enumitem}
\usepackage[T1]{fontenc}
\usepackage{listings}
\usepackage{array}
\usepackage{rotating}
\usepackage{pdflscape}
\usepackage{tikz,pgfplots}
\usepackage{pgfplots}
\usepackage{pgfplotstable}
\pgfplotsset{compat=1.18}
\usepgfplotslibrary{statistics}
\usepackage{comment}
\usepackage{soul}

\newcolumntype{P}[1]{>{\centering\arraybackslash}p{#1}}

\def\tsc#1{\csdef{#1}{\textsc{\lowercase{#1}}\xspace}}
\tsc{WGM}
\tsc{QE}

\begin{document}
\let\WriteBookmarks\relax
\def\floatpagepagefraction{1}
\def\textpagefraction{.001}
\let\printorcid\relax

\newboolean{showcomments}
\setboolean{showcomments}{true} 
\ifthenelse{\boolean{showcomments}}
{\newcommand{\nb}[2]{
		\fcolorbox{black}{yellow}{\bfseries\sffamily\scriptsize#1}
		{\sf\small$\blacktriangleright$\textit{#2}$\blacktriangleleft$}
	}
	\newcommand{\version}{\emph{\scriptsize$-$working$-$}}
}
{\newcommand{\nb}[2]{}
	\newcommand{\version}{}
}

\newcommand\jialong[1]{\nb{Jialong}{#1}}
\newcommand{\methodname}{\textsc{SRTS}}
\renewcommand\cellalign{tr}
\setlength{\tabcolsep}{2pt}
\renewcommand{\arraystretch}{0.95}

\shorttitle{Human Oversight Requirements Perspective in GenAI-enabled Software}

\shortauthors{Z. Mao et al.}  

\title [mode = title]{Towards Requirements Engineering for GenAI-Enabled Software: Bridging Responsibility Gaps through Human Oversight Requirements}

\author[author1]{Zhenyu Mao}
\ead{zhenyumao2-c@my.cityu.edu.hk}

\author[author1]{Jacky Keung}
\ead{Jacky.Keung@cityu.edu.hk}

\author[author1]{Yicheng Sun}
\ead{yicsun2-c@my.cityu.edu.hk}

\author[author1]{Yifei Wang}
\ead{ywang4748-c@my.cityu.edu.hk}

\author[author1]{Shuo Liu}
\ead{sliu273-c@my.cityu.edu.hk}
\cormark[1]
\cortext[cor1]{Corresponding author}

\author[author2]{Jialong Li}
\ead{lijialong@fuji.waseda.jp}

\affiliation[author1]{organization={Department of Computer Science, City University of Hong Kong},
            city={Kowloon},
            country={Hong Kong SAR}}

\affiliation[author2]{organization={Waseda Institute for Advanced Study, Waseda University},     
            city={Tokyo},        
            country={Japan}}


\begin{abstract}
\noindent \textbf{Context}:
Responsibility gaps, long-recognized challenges in socio-technical systems where accountability becomes diffuse or ambiguous, have become increasingly pronounced in GenAI-enabled software.
The generative and adaptive nature of such systems complicates how human oversight and responsibility are specified, delegated, and traced.
Existing requirements engineering approaches remain limited in addressing these phenomena, revealing conceptual, methodological, and artifact-level research gaps in analyzing responsibility gaps from a human oversight requirements perspective.

\noindent \textbf{Objective}:
This study aims to analyze these research gaps in the context of GenAI-enabled software systems.
It seeks to establish a coherent perspective for a systematic analysis of responsibility gaps from a human oversight requirements standpoint, encompassing how these responsibility gaps should be conceptualized, identified, and represented throughout the requirements engineering process.

\noindent \textbf{Methods}:
The proposed design methodology is structured across three analytical layers.
At the conceptualization layer, it establishes a conceptual framing that defines the key elements of responsibility across the human and system dimensions and explains how potential responsibility gaps emerge from their interactions.
At the methodological layer, it introduces a deductive pipeline for identifying responsibility gaps by analyzing interactions between these dimensions and deriving corresponding oversight requirements within established requirements engineering frameworks.
At the artifact layer, it formalizes the results in a Deductive Backbone Table, a reusable representation that traces the reasoning path from responsibility gaps identification to human oversight requirements derivation.

\noindent \textbf{Results}:
A user study compared the proposed methodology with a baseline goal-oriented requirements engineering approach.
Participant-perception results indicated generally favorable views of the enhanced approach across the evaluated dimensions, while expert ratings of the generated requirement artifacts showed higher mean scores for six of the seven artifact-quality items, with the strongest descriptive improvements observed for consistency and representational traceability.

\noindent \textbf{Conclusion}:
The study proposes a methodology for understanding, identifying, and documenting responsibility gaps in GenAI-enabled software from a human oversight requirements perspective.
The results provide initial evidence that it can support the derivation of human oversight requirements from identified responsibility gaps.
However, given the student-dominated sample and the absence of validation in real industrial projects, the generalizability of the findings requires further investigation.

\end{abstract}





\maketitle
\section{Introduction}
\label{sec:introduction}

Generative AI (GenAI) is increasingly embedded in software systems that integrate generative models as active components of reasoning, interaction, and decision support, forming a new class of GenAI-enabled software systems.
Such systems underpin diverse domains and follow recurring design patterns: prompt-based exchanges shape educational and writing assistants \cite{nikolic2024prompt}, retrieval-grounded synthesis informs clinical and research tools \cite{ke2025retrieval}, tool-mediated reasoning supports legal and analytical applications \cite{collenette2023explainable}, and agentic workflows enable autonomous operations in financial and operational platforms \cite{okpala2025agentic}.
Within these systems, humans occupy evolving roles, such as operators, evaluators, supervisors, or auditors, each engaging GenAI components with different levels of initiative and awareness, consistent with findings from related Human--Computer Interaction (HCI) research \cite{tsamados2025human, parasuraman2000model}.
As this new class of software systems matures, it extends human capability and redefines the distribution of control, insight, and responsibility between humans and GenAI.

As their autonomy deepens, they increasingly blur established lines of accountability in digital decision processes.
When generative models produce outputs that shape important outcomes, accountability becomes diffuse as developers cannot see how models reach conclusions, users cannot fully control their behavior, and organizations often struggle to trace decisions after deployment.
This misalignment between assigned responsibility and actual control constitutes a responsibility gap, a condition in which no human agent can be appropriately held to account for an AI system’s actions or consequences \cite{santoni2021four, naumov2025responsibility}.
From a requirements engineering (RE) perspective, this can be viewed as a set of potential responsibility risks, including incomplete, ambiguous, conflicting, or untraceable responsibilities.

To mitigate such risks, accountability must remain anchored to at least one agent who retains both decisive control and informed understanding of system outcomes \cite{naumov2025responsibility}.
In GenAI-enabled software, generative models can influence results but lack epistemic awareness, and therefore cannot satisfy this condition.
Only humans possess the capacity for both decisive and informed control, provided that system design preserves their authority and understanding.
This recognition shifts the focus to RE for GenAI-enabled software, in which accountability relations are operationalized through human oversight mechanisms such as human roles, controls, and decision authorities \cite{eu2024aiact}.
Therefore, from an RE perspective, human oversight can be modeled as a structured set of requirements that specify where human judgment applies, how decisions are reviewed or overridden, and how accountability is maintained.

However, this perspective, treating human oversight as an explicit category in RE, remains underexplored.
Previous research in RE has developed mature perspectives for domains such as privacy \cite{anton2004requirements}, security \cite{mead2005security}, safety \cite{kelly2004goal}, and ethics \cite{guizzardi2023ontology}, where requirement types and derivation methods are well formalized, while comparable foundations for human oversight are still lacking.
When examined through the lens of ISO/IEC/IEEE 29148:2018 \cite{iso29148}, which specifies requirement quality attributes such as completeness, consistency, traceability, feasibility, reusability, and verifiability, existing RE processes for GenAI-enabled software reveal systematic deficiencies.
The absence of these attributes reveals three corresponding research gaps in conceptualization, methodology, and artifact construction.

\begin{table*}[h]
\centering
\caption{Mapping of research gaps and corresponding contributions against ISO/IEC/IEEE~29148 requirement quality attributes}
\label{tab:gaps_contributions_attributes}
\begin{tabular}{|P{2cm}|P{2.5cm}|P{5cm}|P{7cm}|}
\hline
\textbf{Research Gap} & \textbf{ISO~29148 Attribute(s)} & \textbf{Underlying Cause} & \textbf{Contribution in This Work} \\ \hline
Conceptual & Completeness, Consistency & Human oversight lacks a coherent conceptual foundation within the human--GenAI interaction. & Establish a conceptual framing that defines key elements concerning responsibility across the human and system dimensions. \\ \hline
Methodological & Process Traceability, Feasibility & No pipeline connects conceptual oversight recognition to established RE processes. & Propose a deductive pipeline that analyzes responsibility gaps by linking system patterns and human roles, and derives oversight requirements. \\ \hline
Artifact & Representational Traceability, Conformance & No standardized artifact captures or links oversight requirements throughout the RE workflow. & Introduce Deductive Backbone Table as a formal, reusable structure for documenting the reasoning trace from conceptual elements and responsibility gaps to the derived oversight requirements. \\ \hline
\end{tabular}
\end{table*}

The first gap is conceptual.
Despite the growing emphasis on human oversight in AI governance frameworks such as the EU AI Act \cite{eu2024aiact}, ISO 42001 \cite{iso42001}, and the NIST AI RMF \cite{nist2023rmf}, RE still lacks a coherent conceptual foundation for representing oversight as a distinct requirement category alongside privacy, security, safety, and ethics. 
At a foundational level, the field lacks a shared understanding of the positioning of human oversight within the human–GenAI interaction, including its scope, boundaries, and intended function in decision-making and accountability.
At a definitional level, RE research and standards provide no formal descriptions of how both the human and the system dimensions of GenAI-enabled software should be represented in oversight terms.
This conceptual gap prevents RE from ensuring the completeness and consistency of oversight specifications as prescribed by the standard, leaving accountability relations between the human and system dimensions of GenAI-enabled software acknowledged in principle but lacking formal representation.

The second gap is methodological.
Even where human oversight is acknowledged as a desirable property, RE lacks systematic methods for deriving oversight requirements across the software lifecycle.
Existing general-purpose RE frameworks, such as goal-oriented \cite{van2001goal}, scenario-based \cite{sutcliffe1998scenario}, and aspect-oriented approaches \cite{rashid2002early}, provide robust mechanisms for elaborating functional and non-functional requirements once they are defined, but they lack a complementary methodological bridge that connects the conceptual identification of oversight needs with existing RE processes.
Designing this bridge is further complicated by the inherent subjectivity of human analysis, as consistently observed in empirical studies \cite{hidellaarachchi2021effects, hidellaarachchi2023influence}, which affects how oversight is recognized, represented, and refined across contexts.
This methodological gap prevents RE from satisfying the process traceability and feasibility criteria in the standard, leaving oversight requirements disconnected from established RE processes.

The third gap is artifact-related.
Currently, RE lacks a formalized representational structure for documenting and analyzing human oversight requirements in GenAI-enabled software \cite{zave1997four}.
Existing artifacts, such as goal models, use-case templates, or quality attribute taxonomies, capture security, safety, or ethical concerns but provide no consistent means of expressing how human oversight emerges from interactions between system behaviors and human roles.
What is missing is an intermediate representational layer between conceptual oversight concerns and implementation artifacts, one that can encode how oversight relations evolve across design and documentation.
In current practice, such information is scattered across informal documents or embedded implicitly within design discussions, hindering traceability, comparison, and reuse.
This artifact gap prevents RE from satisfying the conformance and representational traceability expectations of the standard, leaving oversight requirements weakly structured and prone to loss across documentation.

To address these gaps, this research proposes a design methodology that formalizes human oversight requirements in GenAI-enabled software.
The key idea is to identify potential responsibility gaps by aligning system-side patterns with human-side roles through a deductive backbone, and to derive corresponding oversight requirements by integrating these analyses into existing RE frameworks.
The methodology is structured across three analytical layers: a conceptualization layer that defines key elements of responsibilities, a methodological layer that operationalizes their interaction through a deductive pipeline, and an artifact layer that represents the results in a formal, reusable documentation structure.
Accordingly, the contributions of this research correspond directly to the three identified research gaps and their associated ISO/IEC/IEEE 29148:2018 requirement quality attributes: completeness and consistency for conceptualization, process traceability and feasibility for methodology, and representational traceability and conformance for artifact representation:
\begin{itemize}
    \item \textbf{Conceptualization contribution}: We establish human oversight as an explicit requirement-level concern in GenAI-enabled software by defining the core responsibility-related elements required for oversight analysis, including system patterns, human roles, and responsibility gaps. This contribution supports completeness by specifying the essential elements that RE analysts should consider when analyzing oversight, and supports consistency by defining the relationships among system-side factors, human-side factors, and the responsibility gaps that emerge from their interaction.
    \item \textbf{Methodological contribution}: We develop a Backbone-Anchored Deductive Pipeline that operationalizes responsibility-gap analysis. The pipeline guides analysts from system pattern analysis and human role analysis to responsibility gap identification and human oversight requirement derivation within established RE frameworks. This contribution supports process traceability by making the reasoning path from analysis inputs to derived requirements explicit, and supports feasibility by integrating the proposed analysis with existing RE processes.
    \item \textbf{Artifact contribution}: We introduce the Deductive Backbone Table as a reusable RE artifact for documenting responsibility-gap analysis and human oversight requirement derivation. The table records the relationships among system patterns, human roles, responsibility gaps, and derived oversight requirements. This contribution supports representational traceability by preserving these links in a structured artifact, and supports conformance by aligning the documentation with the proposed deductive backbone and established RE quality expectations.
\end{itemize}
Table~\ref{tab:gaps_contributions_attributes} summarizes the alignment among the research gaps, associated requirement quality attributes, and corresponding contributions.

The remainder of this paper is organized as follows.
Section \ref{sec:background} reviews the background and related work.
Section \ref{sec:proposal} introduces the proposed methodology, including conceptualization, methodological, and artifact layers.
Section \ref{sec:user_study} describes the user study design and results.
Finally, Section \ref{sec:conclusion} concludes the paper and outlines future research directions.
\section{Background and Related Work}
\label{sec:background}

\subsection{Requirements Engineering Foundations}

Requirements Engineering (RE) provides the methodological backbone of software development, encompassing the elicitation, specification, validation, and management of system requirements to ensure that software systems satisfy stakeholder needs and conform to quality expectations \cite{zave1997four, nuseibeh2000requirements}.
Among the most influential are goal-oriented approaches \cite{van2001goal, yu1997towards, dardenne1993goal}, which derive requirements from stakeholder intentions and system objectives,  scenario-based approaches \cite{sutcliffe1998scenario, carroll2003making, rosson2002usability}, which capture behavioral expectations through user stories and interaction contexts, and aspect-oriented approaches \cite{rashid2002early, rashid2003modularisation, moreira2002crosscutting}, which manage cross-cutting concerns such as security, privacy, or usability that affect multiple system components.
Together, these approaches provide analytical structures for decomposing and organizing complex requirement sets across both functional and non-functional dimensions \cite{zave1997four, mylopoulos1992representing}.

To ensure quality and consistency across these practices, RE is supported by a family of international standards that define processes, artifacts, and evaluation criteria throughout the software and systems lifecycle.
Among the most influential are ISO/IEC/IEEE 29148:2018 \cite{iso29148}, which specifies requirement quality attributes such as completeness, consistency, traceability, feasibility, verifiability, and reusability, ISO/IEC 12207:2017 \cite{iso12207} and ISO/IEC/IEEE 15288:2023 \cite{iso15288}, which define lifecycle processes for software and systems engineering, and ISO/IEC 25002:2024 \cite{iso25002}, which establishes quality models that guide the specification and assessment of non-functional requirements.

Despite its methodological maturity, RE remains constrained by the inherent subjectivity of human interpretation as requirements elicitation and analysis depend on how analysts perceive problems, frame system goals, and negotiate meanings with stakeholders.
Empirical studies show that even experienced engineers vary in how they identify, classify, and prioritize requirements, leading to inconsistencies in completeness and traceability \cite{hidellaarachchi2021effects, hidellaarachchi2023influence}.
This variability does not weaken the discipline but emphasizes its human-centered nature, in which the success of RE depends on both the professional judgment of analysts and the methodological support (e.g., standardizing reasoning patterns) that mitigates subjectivity in requirements analysis \cite{boehm2011software, hidellaarachchi2021effects, zhang2024adaptive}.

\subsection{Requirements Categories}

RE has long differentiated specialized requirement categories to address distinct classes of system quality, risk, and domain concern.
Such categories, exemplified by safety, security, privacy, and ethics, often encompass both functional and non-functional aspects, reflecting the multifaceted nature of socio-technical systems.
Each has matured into a dedicated subfield with its own concepts, modeling artifacts, and assurance practices, illustrating how RE evolves to formalize emerging contextual concerns into systematic requirement types as technology and society co-evolve.

Safety requirements address the need to prevent hazardous system behavior in domains where failures can have catastrophic consequences, such as avionics, medical devices, and industrial automation.
To manage such risks, structured approaches were developed within RE practice, ranging from goal-based safety cases and formal hazard analyses to system-theoretic frameworks like STAMP, that link functional behavior, control structures, and human decision processes to risk mitigation \cite{kelly2004systematic, hollnagel2016barriers, leveson2016engineering}.
As systems became increasingly exposed to networked environments, security requirements emerged as a distinct category addressing protection from intentional misuse, data breaches, and unauthorized access.
To support systematic analysis of these risks, structured frameworks, such as SQUARE and abuse cases were developed, integrating threat modeling, stakeholder analysis, and requirements prioritization into early RE stages \cite{mead2005security, haley2008security, mcdermott1999using}.
In parallel, privacy requirements gained recognition as distinct from security by emphasizing principles such as purpose limitation, data minimization, and informed consent.
In response, approaches such as goal-oriented modeling, privacy threat analysis, and regulatory rule extraction have been developed to capture information flows, model consent and disclosure relationships, and translate legal obligations into traceable requirements \cite{anton2002analyzing, wuyts2014empirical, breaux2008analyzing}.
More recently, ethical requirements have expanded RE to address challenges of ensuring fairness, transparency, and accountability in algorithmic and data-driven systems, to ensure that automated reasoning and decision processes remain aligned with human values throughout the system lifecycle.
To support this alignment, approaches such as ontology-based ethical modeling, value-sensitive design, and responsible AI engineering frameworks have been proposed to formalize moral principles and integrate them into verifiable requirement specifications \cite{guizzardi2023ontology, aydemir2018roadmap, theodorou2020towards}.

These requirement categories show how RE has historically incorporated emerging socio-technical concerns by developing dedicated concepts, analysis methods, and documentation artifacts.
Closely related streams, including safety engineering, accountability frameworks, and assurance case modeling, provide further foundations for reasoning about risk, responsibility, and evidence.

Safety engineering focuses on identifying hazards, deriving safety constraints, and demonstrating risk mitigation through control-oriented analysis and safety arguments \cite{leveson2004new}.
For example, STPA-based approaches have been used to derive software safety requirements from system-level safety analysis \cite{abdulkhaleq2015comprehensive}.
However, safety engineering primarily targets hazardous system behavior and risk reduction, whereas this study focuses on requirements for preserving accountable human oversight when responsibility for GenAI-enabled Software outcomes becomes ambiguous, incomplete, or difficult to trace.
Accountability frameworks clarify responsible AI principles, role obligations, and governance expectations across the AI lifecycle \cite{mittelstadt2019principles}.
For example, internal algorithmic auditing frameworks provide lifecycle-oriented documentation and review mechanisms for organizational accountability \cite{raji2020closing}.
However, such frameworks often remain at the governance or audit-process level, whereas this study provides an RE-oriented method for deriving concrete oversight requirements.
Assurance case modeling focuses on constructing structured arguments and evidence to justify confidence in system properties such as safety or security \cite{bloomfield2009safety}.
For example, model-based assurance approaches use notations and metamodels such as GSN or SACM to organize claims, arguments, and evidence \cite{wei2019model}.
However, assurance cases typically support the justification of already identified assurance properties, whereas this study addresses the earlier RE task of identifying responsibility gaps and deriving the oversight requirements to be specified.

\subsection{Responsibility Gaps}

A responsibility gap occurs when it becomes indeterminate who can justifiably be held morally, legally, or practically responsible for the behavior or consequences of an autonomous system.
The term denotes a condition in which standard criteria for attributing responsibility, such as knowledge, control, and causal connection, no longer hold \cite{matthias2004responsibility}. 
Subsequent analyses broaden the concept from individual culpability to systemic and institutional dimensions, distinguishing four interrelated forms, including culpability, moral-accountability, public-accountability, and active-responsibility gaps \cite{santoni2021four}.
Collectively, these forms reflect how technical and organizational dynamics interact to erode control and accountability, as autonomy and opacity undermine decision traceability while distributed agency and institutional fragmentation blur the allocation of authority and responsibility \cite{coeckelbergh2020ai, johnson2019ai, konigs2022artificial}.

Responsibility gaps in GenAI-enabled software systems arise from the interplay between generative model behavior and human oversight structures, producing proxy gaps when generative agents act as representational proxies for human users yet operate with partial autonomy that weakens both knowledge and control \cite{constantinescu2025genai}.
Such systems exemplify proxy gaps, where accountability is dispersed across the socio-technical arrangement where epistemic gaps limit users’ understanding of how outputs are produced and control gaps weaken their capacity to influence outcomes.
The proxy gap demonstrates how classical responsibility gaps resurface in generative settings, where model-mediated delegation and fragmented human roles obscure agency and accountability.
Although earlier discussions address responsibility gaps in autonomous and decision-making systems, analyses of GenAI-enabled software remain scarce.
The proxy gap case represents an initial exploration of these issues within GenAI, offering conceptual insight but not yet a structured approach to their systematic understanding.

\subsection{GenAI-enabled Software Systems}

The growing integration of GenAI into software architectures has given rise to a distinct class of GenAI-enabled software systems.
These systems differ from GenAI systems, which primarily concern the design and operation of generative models themselves.
They instead encompass the broader socio-technical environments in which such models are embedded as functional components that mediate reasoning, interaction, and decision processes \cite{vandeputte2025foundational}.
Within these systems, generative components operate as adaptive, co-creative subsystems, producing context-dependent outputs that evolve through probabilistic reasoning and user feedback.
Recent studies show that LLMs can synergize with automated machine learning workflows and support runtime reasoning, adaptation planning, and system evolution, while also raising open challenges related to reliability, controllability, and human oversight \cite{xu2024large, li2024generative}.
This integration transforms conventional software architectures into dynamic assemblages of human and machine agency, producing synthesized rather than deterministic outputs.

The design of GenAI-enabled software is fundamentally an architectural problem, structuring how generative components are contained, coordinated, and made observable within complex environments \cite{vandeputte2025foundational}, and it requires concurrent attention to control and transparency.
Ensuring control involves architectural mechanisms such as modular containment, orchestration, and governance interfaces that constrain generative behavior and sustain system reliability while preserving adaptivity \cite{esposito2025generative}.
Likewise, ensuring transparency depends on exposing reasoning, provenance, and uncertainty through interpretable outputs and traceable process representations that support accountable human oversight \cite{cheng2024generative}.

\subsection{Human Oversight}

Human oversight refers to the structured involvement of human agents in monitoring, guiding, and intervening in the behavior of autonomous or generative systems to ensure that outcomes remain aligned with ethical, legal, and organizational expectations \cite{europeancommission2024aiact}.
In the context of GenAI-enabled software, such oversight serves as a key mechanism for closing responsibility gaps by ensuring that humans, rather than systems, retain decisive control over critical decisions and informed understanding of system outcomes. 
Oversight transforms accountability from a reactive attribution process into a proactive design objective, embedding human judgment and intervention pathways throughout the system lifecycle to maintain responsibility and trust when generative behavior becomes opaque or unpredictable \cite{sterz2024quest}.
Yet, despite its centrality to accountability, human oversight remains underdeveloped as a requirement category in RE and needs to be formalized to systematically institutionalize the human role within adaptive and semi-autonomous systems.

To operationalize such oversight within GenAI-enabled systems, its effectiveness depends on how authority and interaction are structured within human–AI collaboration, as decades of HAI research consistently identify decision control and communicative transparency as the dual foundations of successful human–automation coordination \cite{schaefer2016meta, amershi2019guidelines}.
Maintaining authority requires explicit boundaries of decision rights, escalation pathways, and override capacities that enable humans to retain meaningful control and accountability as automation levels increase \cite{parasuraman2000model}.
Ensuring interaction relies on transparent communication and adaptive feedback mechanisms that allow humans to interpret, calibrate trust in, and effectively collaborate with automated systems \cite{lee2004trust}. 
\section{Proposed Methodology}
\label{sec:proposal}

\subsection{Methodology Overview}
\label{sec:overview}

\begin{figure*}[hbtp]
    \centering
    \includegraphics[width=170mm]{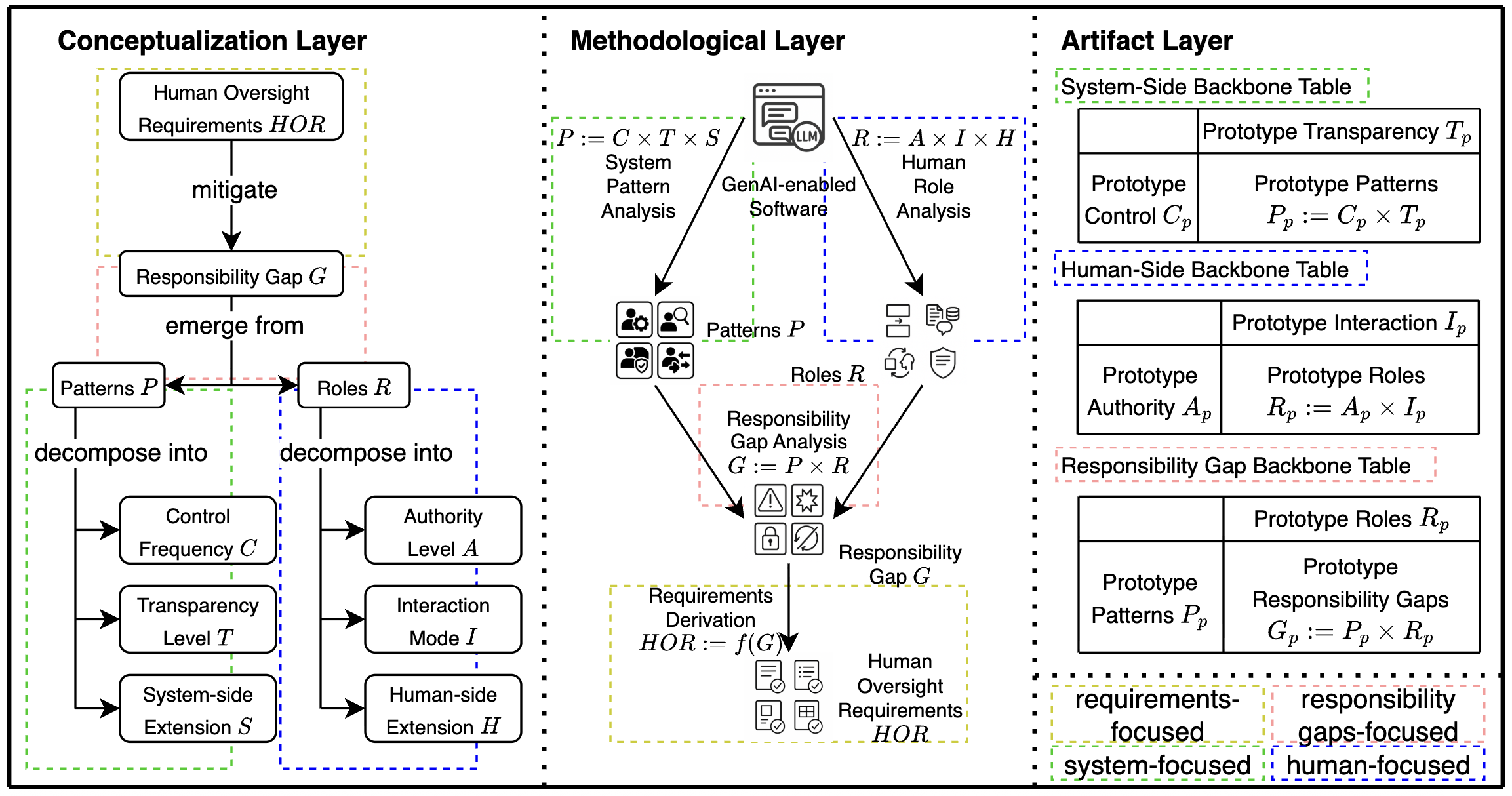}
    \caption{Methodology Overview}
    \label{fig:methodology}
\end{figure*}

This section presents the proposed design methodology for analyzing human oversight requirements in GenAI-enabled software through the systematic examination of responsibility gaps.
The methodology provides a structured pathway that defines the conceptual basis of responsibility and explains how responsibility gaps emerge across the human and system dimensions.
It then derives corresponding oversight requirements through a deductive analysis of these gaps and documents the reasoning trace from responsibility gap identification to oversight requirement derivation in a formal, reusable representation.
It comprises three interconnected layers, conceptualization, methodological, and artifact, each addressing a distinct aspect of the oversight requirements lifecycle, as shown in Figure \ref{fig:methodology}.

The conceptualization layer defines the theoretical foundations required for deriving human oversight requirements ($HOR$).
It introduces the conceptual structures through which responsibility is analyzed, including patterns ($P$) derived from the system dimension, control frequency ($C$), transparency level ($T$), and system-side extensions ($S$), and roles ($R$) derived from the human dimension, authority level ($A$), interaction mode ($I$), and human-side extensions ($H$).
The methodological layer establishes a Backbone-Anchored Deductive Pipeline that links these conceptual elements, analyzes the resulting responsibility gaps ($G$), and derives the corresponding human oversight requirements ($HOR$).
The artifact layer introduces the Deductive Backbone Table as a formal representational structure that can be instantiated as multiple tables to ensure traceability and conformance by documenting the process of identifying responsibility gaps and deriving human oversight requirements ($HOR$) across roles and system patterns.
The notations employed in the methodology are summarized in Table~\ref{tab:notations}.

\begin{table*}[h]
\centering
\caption{Notations used in the methodology}
\label{tab:notations}
\begin{tabular}{|P{1cm}|P{15.5cm}|}
\hline
\textbf{Symbol} & \textbf{Concept Represented} \\ \hline
$C$ & Control frequency --- backbone system factor indicating how often humans can intervene or adjust system behavior.\\ \hline
$T$ & Transparency level --- backbone system factor indicating the degree of interpretability or visibility of the reasoning.\\ \hline
$S$ & System-side extensions --- analyst-defined factors that extend or refine the deductive backbone $C\times T$. \\ \hline
$A$ & Authority level --- backbone human factor defining the degree of human decision power in system operations.\\ \hline
$I$ & Interaction mode --- backbone human factor defining how humans engage with the system.\\ \hline
$H$ & Human-side extensions --- analyst-defined factors that extend or refine the deductive backbone $A\times I$. \\ \hline
$P$ & Patterns --- characteristic configurations of generative system behavior relevant to oversight.\\ \hline
$R$ & Roles --- characteristic configurations of human authority, interaction, and contextual factors relevant to oversight.\\ \hline
$G$ & Responsibility gaps --- potential misalignments between system patterns and human roles regarding responsibility.\\ \hline
$HOR$ & Human oversight requirements --- specifications defining how oversight structures and processes should be designed to realign responsibility and ensure accountability.\\ \hline
\end{tabular}
\end{table*}

\subsection{Conceptualization Layer}
\label{sec:conceptualization}

Following the requirement quality attributes discussed in ISO/IEC/IEEE 29148:2018, the conceptualization layer supports completeness, interpreted here as covering the necessary responsibility-related elements for specifying human oversight requirements, by explicitly defining system-side, human-side, and gap-related elements as the required analytical scope.
It also supports consistency, interpreted here as defining these elements and their relationships without contradiction or incompatible interpretation.
This is achieved by organizing them through a coherent deductive structure in which system patterns and human roles jointly determine responsibility gaps, and responsibility gaps motivate human oversight requirements.

The conceptualization follows a top-down logic, beginning with the goal of defining human oversight requirements ($HOR$) and tracing the conceptual dependencies that make their specification possible.
These requirements define the conditions under which oversight must be established and maintained so that responsibility and accountability remain explicit throughout the lifecycle of GenAI-enabled systems.
They articulate the structural and procedural means through which humans monitor, guide, and intervene in generative behavior to prevent the loss of control or traceability.
Yet, even when such mechanisms are in place, responsibility may still become diffuse or uncertain within complex GenAI-enabled systems.

Responsibility gaps ($G$) refer to conditions where the allocation of responsibility between human agents and GenAI components becomes ambiguous or incomplete.
Such gaps arise when no human role can be clearly identified as accountable for a system behavior, or when existing oversight capacity is insufficient to influence or interpret generative outcomes.
Within the methodology, identifying $G$ is a critical analytical step for defining $HOR$, as each oversight requirement addresses a specific gap by restoring a clear linkage between human authority and system behavior.
Understanding how these gaps emerge requires examining the underlying structures through which human and system factors interact.

System patterns ($P$) and human roles ($R$) provide the structural foundation for understanding how responsibility is distributed within GenAI-enabled software systems.
Patterns and roles represent characteristic forms of generative system behavior and human engagement, together defining a conceptual structure with multiple dimensions that describe how human and system factors interact to shape responsibility.
On the system side, relevant dimensions include control frequency ($C$), which indicates how often humans can intervene or adjust generative processes, and transparency level ($T$), which indicates how much of the system’s reasoning and uncertainty is made interpretable, along with other architectural or operational factors that influence these properties.
On the human side, key dimensions include authority level ($A$), which defines the degree of decision power retained by humans, and interaction modes ($I$), which define how humans engage with and influence the system, complemented by contextual or organizational aspects that further shape oversight practices.
Responsibility gaps ($G$) arise when specific patterns and roles are misaligned across these dimensions, creating conditions where responsibility cannot be clearly assigned or enacted.
These dimensions therefore outline the conceptual coordinates for understanding the alignment between system behaviors and human oversight, forming the basis for identifying $G$ and deriving corresponding human oversight requirements ($HOR$).
\subsection{Methodological Layer}
\label{sec:methodology}

\subsubsection{Methodological Layer Overview}

The methodological layer supports process traceability, interpreted here as making the reasoning path from analysis inputs to derived requirements explicit and justifiable.
It is achieved by decomposing the analysis into ordered steps from system pattern analysis and human role analysis to responsibility gap identification and requirement derivation.
It also supports feasibility, interpreted here as enabling the method to be applied within existing RE activities and project constraints, by embedding responsibility-gap analysis into established RE frameworks rather than requiring a separate requirements process.

Building on the conceptual dimensions defined in Section~\ref{sec:conceptualization}, the methodological layer introduces the Backbone-Anchored Deductive Pipeline, which identifies responsibility gaps and operationalizes established RE principles in a deductive style for GenAI-enabled software systems.
The pipeline combines a stable deductive backbone with analyst-driven extensions, ensuring both methodological rigor and contextual adaptability.

The backbone represents the analytical foundation formed by the intersections of essential dimensions on the system and human sides.
These intersections, expressed as $C \times T$ and $A \times I$, define the core analytical space from which responsibility alignment is examined.
The backbone provides a consistent reasoning structure that anchors the analysis, while its extensibility allows analysts to introduce additional system-side ($S$) or human-side ($H$) factors to capture contextual, organizational, or domain-specific oversight concerns.
It should be noted that each dimension can encompass multiple layers, for example, transparency (\(T\)) may include informational, procedural, and interpretive facets, but is currently simplified as a single analytical attribute to maintain a concise and tractable backbone.
These designs ensure that the pipeline remains both principled and flexible across application contexts.
The following subsections outline the operation of the Backbone-Anchored Deductive Pipeline and illustrate its reasoning flow through a concrete scenario. 

\textbf{Illustrative scenario.} 
Consider a GenAI-enabled legal contract review assistant deployed in a corporate law firm.
Once a draft contract is uploaded, the assistant analyzes each clause, flags potential risks, and proposes revisions.
Human professionals remain responsible for final approval, where a paralegal reviews the model’s suggestions before submitting them to a senior lawyer, who oversees the overall review process.
In practice, the assistant conducts most analyses autonomously, with limited opportunities for real-time adjustment or inspection of its reasoning.
Its recommendations are often plausible but difficult to trace back to concrete legal rules or precedents. 
Because contract review involves confidentiality and liability, even minor misjudgments can carry serious consequences.

This situation captures the type of tension the methodology aims to analyze, where automation amplifies efficiency but also obscures how authority, interpretability, and accountability are distributed between human and generative agents.
We use this scenario throughout the following pipeline steps to demonstrate how the proposed methodology operates in practice.

\subsubsection{Step 1 --- System Pattern Analysis}

The first step conceptualizes the GenAI-enabled system through a pattern-based abstraction inspired by software design patterns, which describe reusable solutions to recurring design problems.
This study adapts that idea for responsibility analysis: the term system pattern refers not to a reusable architectural or implementation solution in the conventional design-pattern sense, but to a recurring responsibility-relevant configuration of GenAI-enabled system behavior.
It captures characteristic configurations of generative autonomy and human controllability that determine how responsibility is technologically distributed.
Formally, it is expressed as:
\[
P := C \times T \times S
\]
where $C$, $T$, and $S$ denote the control, transparency, and system-side extensional dimensions introduced earlier. 

In practical application, $C$ may be examined using concrete intervention points such as pre-generation parameter setting, mid-generation steering, or post-generation approval, as well as the extent to which these interventions influence system behavior. 
$T$ can be analyzed through informational, procedural, and interpretive aspects, for example the visibility of reasoning steps, the provenance of evidence, and the comprehensibility of outputs for human reviewers. 
$S$ allows analysts to capture domain-specific or architectural factors, such as uncertainty representation, autonomy modality, or data sensitivity, when these materially affect oversight.
In use, these dimensions may be instantiated differently across GenAI system types. 
For example, a prompt-driven LLM may place most controllability in the pre- and post-generation phases, whereas an agentic system operating through multi-step tool use shifts analytical attention towards monitoring intermediate actions. 

\textbf{Illustrative scenario.} 
In the legal contract review assistant, the GenAI component operates largely autonomously once a contract is submitted for analysis. 
It performs clause-level evaluations and generates revision suggestions without intermediate human steering, indicating a low control frequency ($C_L$). 
Its reasoning for identifying risks or proposing changes is not fully interpretable to human reviewers, reflecting a low transparency level ($T_L$). 
For demonstration, the system-side extensional dimension $S$ is instantiated with a single factor, \textit{sensitivity}, characterized as high (\textit{sensitivity\_high}) due to the potential legal and financial implications of misinterpretation. 
The resulting pattern instance is:
\[
p = (C_L, T_L, \{sensitivity\_high\})
\]
This configuration represents a pattern of autonomous generation under high contractual sensitivity, where the system’s outputs directly influence legally binding content but are produced with minimal human control and limited interpretability.

\subsubsection{Step 2 --- Human Role Analysis}

The second step examines the human layer of responsibility by identifying the roles that interact with the generative system. 
Analogous to system patterns, human roles $R$ are represented as structured configurations that describe how humans exercise control, judgment, and accountability within the system context.
Formally, this relationship is expressed as:
\[
R := A \times I \times H
\]
where $A$, $I$, and $H$ denote the authority, interaction, and human-side extensional dimensions, respectively.

In practical application, $A$ and $I$ can be instantiated by examining the concrete oversight privileges available to each role, such as who may approve, escalate, or override outputs, and whether the role engages in control, validation, monitoring, or corrective actions.
$H$ further allows analysts to incorporate contextual factors, including expertise, workload, or regulatory obligations, that shape how reliably these responsibilities can be carried out in practice.

\textbf{Illustrative scenario.}  
In the legal contract review assistant, analysts begin by determining how authority and interaction are distributed among the human agents involved in the review workflow. 
Because neither actor directly operates or trains the model, operational authority is excluded.  
Likewise, the process is not purely retrospective, so audit-oriented authority is not applicable.  
Both human participants supervise the model’s outputs and remain accountable for the quality of resulting contracts; thus, the appropriate level of authority is supervisory.

The paralegal interacts with the system by directly reviewing clause-level suggestions, approving or modifying them before submission. 
This behavior corresponds to a validation-oriented interaction. 
When combined with supervisory authority, this configuration can be formalized as:
\[
r_{reviewer} = (\{A_{supervisory}\}, \{I_{validation}\}, \emptyset)
\]
which defines the role archetype of a \textit{Reviewer}.

The senior lawyer oversees the entire review process, checking that contracts comply with professional and regulatory standards but seldom intervenes directly in model outputs. 
This engagement represents a monitoring interaction paired with the same supervisory authority, formalized as:
\[
r_{coordinator} = (\{A_{supervisory}\}, \{I_{monitoring}\}, \emptyset)
\]
corresponding to the role archetype of a \textit{Coordinator}.

These two configurations express how authority and interaction combine to form distinct human oversight roles within the system.
Together, they define the human-side structures to be aligned with the system pattern in the next step.

\subsubsection{Step 3 --- Responsibility Gap Analysis}

The third step aligns the system-side and human-side configurations to identify how mismatches in controllability, transparency, or authority give rise to responsibility gaps. 
Each pairing between the system pattern and a human role marks a potential locus of requirement ambiguity, where the intended distribution of control and accountability may not be consistently realized in practice.
Formally, this step is defined as:
\[
G := P \times R
\]
where each element $g = (p, r)$ denotes an interaction between the system operating under pattern $p$ and a human role $r$, through which oversight adequacy and potential gaps can be examined.
In practice, $P$ typically contains only one element, as each GenAI-enabled software system is represented by a single system pattern instance.

\textbf{Illustrative scenario.}  
In the legal contract review assistant, the system pattern derived in Step 1 interacts with the two role configurations defined in Step 2, producing two analytical pairings:
\[
g_1 = (p, r_{\text{reviewer}}), \qquad
g_2 = (p, r_{\text{coordinator}}).
\]

For $g_1$, the paralegal acts as reviewer, evaluating clause-level suggestions that originate from an autonomous, low-transparency process.  
Although formally responsible for approving revisions, the reviewer may not be able to fully interpret the model’s reasoning or trace the evidence supporting each recommendation.
The high sensitivity of the contractual domain can exacerbate this limitation that even minor misinterpretations could lead to financial or legal consequences, yet the reviewer has limited means to verify the model’s judgments beyond surface plausibility.
This configuration exposes an \textbf{accountability gap under high contractual sensitivity}, in which decision authority is nominally assigned but not supported by sufficient interpretability, evidential access, or control over the generative process.

For $g_2$, the senior lawyer functions as coordinator, overseeing the review process and ensuring compliance with organizational and legal standards.
However, because the system operates autonomously with limited transparency, the coordinator has only indirect visibility into how specific risk assessments or clause suggestions are generated.
When sensitivity is high, such as in contracts involving confidentiality, liability, or regulatory constraints, this restricted authority may lead to blurred lines of accountability, as responsibility for the system’s behavior could be distributed yet insufficiently claimed.  
This misalignment gives rise to an \textbf{ownership gap under high contractual sensitivity}, where supervisory awareness exists in principle but lacks formal mechanisms for authority and accountability alignment.

\subsubsection{Step 4 --- Requirements Derivation}

The fourth step derives explicit human oversight requirements from the potential responsibility gaps identified in Step 3.
Existing RE frameworks support this derivation by providing structured mechanisms that link responsibility analysis with conventional requirement modeling practices.
Formally, the derivation process is expressed as:
\[
HOR := f(G)
\]
where $G$ denotes the set of potential responsibility gaps, and $f$ represents the transformation of each gap into framework-specific modeling constructs that yield human oversight requirements as outputs.

Different RE paradigms implement this transformation in distinct ways.  
In goal-oriented requirements engineering, analysis results obtained in previous steps guide the definition or refinement of goals that capture the related oversight concern, which is then analyzed through obstacles and resolved by corresponding human and system requirements.
In scenario-based approaches, each gap gives rise to a dedicated scenario specifying the human–system interactions necessary to prevent or mitigate the gap.  
In aspect-oriented modeling, each gap defines a cross-cutting aspect that enforces properties such as traceability, interpretability, or approval authority across modules.

\textbf{Illustrative scenario.}  
Continuing with the legal contract review assistant, the system pattern, role configurations, and potential gaps identified in Steps 1–3 form the structured inputs for this stage.  
They are integrated through the five reasoning steps of goal-oriented requirements engineering, identify main goals, refine goals, assign agents, find obstacles, and derive requirements, to convert the analytical findings into concrete oversight requirements.

\textit{Step 1: Identify Main Goal.} 
The system pattern indicates autonomous generation under high contractual sensitivity, where outputs directly influence legally binding content but are produced with minimal human control and limited interpretability. 
This profile prioritizes accountability and verifiability: autonomy reduces opportunities for intervention, low transparency limits justificatory evidence, and high sensitivity amplifies the consequences of error. 
Accordingly, the main goal is defined as follows.

\textbf{G1:} Ensure that AI-generated revisions are reviewed and approved through transparent and traceable evidence.

\textit{Step 2: Refine Goals.}  
Using the human roles identified in Step 2 as reference points, the main oversight goal (G1) is refined into role-specific objectives that specify what each actor ensures as part of achieving it.
Each refined goal maintains direct traceability to G1 by indicating both the responsible role and its contribution.

\begin{itemize}
  \item \textbf{G1.1 (G1-r1):} Validate the accuracy, relevance, and legal soundness of each AI-suggested revision, confirming that all proposed changes are supported by adequate reasoning and evidence before approval.
  \item \textbf{G1.2 (G1-r2):} Provide clause-level explanations, confidence indicators, and linked source references that allow the reviewer to verify how each suggestion was derived.
\end{itemize}

\textit{Step 3: Assign Agents.}  
For each refined goal, the corresponding responsible actor is identified to establish clear accountability within the scenario.  
Each mapping preserves the same goal identifiers (G1.1, G1.2) to maintain traceability between analytical and practical representations.

\begin{itemize}
  \item \textbf{G1.1} Paralegal responsible for validating the model’s clause-level revisions before approval.
  \item \textbf{G1.2} The generative contract review assistant that produces revision suggestions and provides supporting evidence for human verification.
\end{itemize}

\textit{Step 4: Find Obstacles.}  
For each refined goal–agent pair, identify the specific conditions that may prevent the responsible actor from achieving its goal.
Each obstacle reflects how the responsibility gaps diagnosed in Step 3 could appear in this scenario, translating structural misalignments into concrete barriers to effective oversight.

\begin{itemize}
  \item \textbf{O1 (G1.1 – r1):} Limited model transparency and lack of accessible reasoning may prevent the reviewer from confidently validating AI-suggested revisions, risking approval of content that cannot be fully justified.
  \item \textbf{O2 (G1.2 – r2):} The system’s autonomous generation process may not include explicit evidence-tracking or explanation mechanisms, restricting its ability to provide clause-level justification for its outputs.
\end{itemize}

\textit{Step 5: Derive Requirements.}  
For each obstacle identified in the previous step, one or more requirements are defined to remove, reduce, or mitigate its impact.  
Each requirement specifies an actionable measure to close the potential accountability or oversight gap, expressed as either a system-side or a human-side commitment.  
System-side requirements describe technical or interface features that enable effective human oversight, while human-side requirements focus on procedural, training, or policy actions that ensure responsible use and supervision.  
Each requirement is labeled for traceability as \textbf{R1s} (system) or \textbf{R1h} (human), corresponding to the related obstacle (\textbf{O1}), and similarly for subsequent pairs.

\begin{itemize}
    \item \textbf{R1s (System):} The generative contract review assistant shall display traceable source references, confidence indicators, and concise reasoning summaries for each suggested revision to support human verification.
    \item \textbf{R1h (Paralegal):} The paralegal responsible for reviewing AI-suggested revisions shall verify that each proposed change is supported by adequate evidence and justification before approving contractual content.
    \item \textbf{R2s (System):} The generative contract review assistant shall maintain detailed logs of reasoning steps and evidence sources for all clause-level analyses to enable oversight of its decision process.
    \item \textbf{R2h (Senior Lawyer):} The senior lawyer overseeing the review process shall conduct periodic audits of the system’s logs to confirm that explanations and evidence meet the firm’s legal and traceability standards.
\end{itemize}

Together, these requirements restore alignment between human and system responsibilities under high contractual sensitivity by ensuring that the Reviewer’s accountability and the Coordinator’s oversight authority are supported.
\subsection{Artifact Layer}
\label{sec:artifact}

\subsubsection{Artifact Layer Overview}

The artifact layer supports representational traceability, interpreted here as preserving explicit links among system patterns, human roles, responsibility gaps, and derived requirements, by recording these elements and their relationships in the Deductive Backbone Table.
It also supports conformance, interpreted here as aligning the documentation with the proposed deductive backbone and established RE quality expectations, by standardizing how analytical inputs, gap reasoning, and derived oversight requirements are documented.

The Deductive Backbone Tables form both the analytical and empirical core of the Backbone-Anchored Deductive Pipeline. 
They have a dual nature in that they function analytically as deductive scaffolds that formalize the reasoning space of responsibility analysis through the intersections of essential dimensions including control ($C$), transparency ($T$), authority ($A$), and interaction ($I$). 
At the same time they operate empirically as synthesized representations of an analysis that describe how autonomy, interpretability, authority, and engagement are distributed within a GenAI-enabled system. 
By corresponding to analytical steps in the Backbone-Anchored Deductive Pipeline, these tables transform the reasoning process of the methodological layer into reusable documentation artifacts.
Each table therefore acts as a bridge between conceptual reasoning and applied evaluation and ensures that responsibility analysis remains both theoretically grounded and operationally interpretable.

The three Backbone Tables together form the pipeline’s full reasoning structure, with the System-Side Backbone Table ($C \times T$) characterizing generative behavior, the Human-Side Backbone Table ($A \times I$) mapping oversight roles, and the Responsibility Gap Backbone Table ($P \times R$ or $C \times T \times A \times I$) integrating both to reveal where misalignments may produce responsibility gaps.
In practical use the Deductive Backbone Tables provide a structured shortcut throughout the analytical process. 
Once the relevant values of the system and human dimensions are determined, the analyst can refer directly to the corresponding table instead of reinterpreting each configuration from the beginning. 
For example, when a configuration such as $C_L \times T_L \times \{sensitivity\_{high}\}$ appears, its meaning can be located directly in the System-Side Backbone Table \ref{tab:ct_backbone} as a case of autonomous generation under a high sensitivity context. 
This approach ensures analytical consistency across all stages of reasoning and improves efficiency since the tables encapsulate the essential reasoning space that supports interpretation and oversight derivation.

\subsubsection{System-Side Backbone Table ($C \times T$)}

\begin{table*}[htbp]
\centering
\caption{System-Side Backbone Table ($C \times T$)}
\label{tab:ct_backbone}
\renewcommand{\arraystretch}{1.25}
\begin{tabular}{|P{3cm}|P{5.5cm}|P{5.5cm}|}
\hline
\textbf{$C \times T$} & \textbf{Transparency High ($T_H$)} & \textbf{Transparency Low ($T_L$)} \\ \hline
\textbf{Control High ($C_H$)} 
& \textbf{Co-generation} — systems where humans and the model collaborate continuously with high interpretability and frequent control, such as interactive design support or adaptive writing assistants. 
& \textbf{Blind steering} — systems with frequent human interventions but limited visibility into the model reasoning, such as real-time recommendation tools with opaque decision updates. \\ \hline
\textbf{Control Low ($C_L$)} 
& \textbf{Review and approve} — systems where the model produces interpretable outputs that humans validate after generation, such as structured content generation or report synthesis systems. 
& \textbf{Autonomous generation} — systems that operate independently with minimal interpretability or control, such as automated summarization or decision-support modules functioning without intermediate feedback. \\ \hline
\end{tabular}
\end{table*}

\begin{table*}[htbp]
\centering
\caption{Human-Side Backbone Table ($A \times I$)}
\label{tab:ai_backbone}
\renewcommand{\arraystretch}{1.25}
\begin{tabular}{|P{2cm}|P{3.2cm}|P{3.2cm}|P{3.2cm}|P{3.2cm}|}
\hline
\textbf{$A \times I$} & \textbf{Active Control ($I_1$)} & \textbf{Approval / Validation ($I_2$)} & \textbf{Monitoring / Auditing ($I_3$)} & \textbf{Corrective / Maintenance ($I_4$)} \\ \hline
\textbf{Operational ($A_1$)} 
& \textbf{Operator} — performs direct control of generative processes through continuous interaction. 
& Rare configuration; validation normally requires independent authority. 
& Limited oversight value; duplicates operational actions. 
& \textbf{Maintainer} — performs adjustments and corrections after generation. \\ \hline
\textbf{Supervisory ($A_2$)} 
& Uncommon; high authority rarely performs low-level control. 
& \textbf{Reviewer} — validates or approves generative outcomes. 
& \textbf{Coordinator} — oversees processes and manages communication among actors.
& Typically delegated; supervisors set corrective policies but seldom act directly. \\ \hline
\textbf{Audit ($A_3$)} 
& Incompatible with audit independence. 
& Conflicts with audit mandate. 
& \textbf{Auditor} — retrospectively evaluates compliance and traceability. 
& Governance-level; links audit outcomes to organizational policy. \\ \hline
\end{tabular}
\end{table*}

The System-Side Backbone Table \ref{tab:ct_backbone} operationalizes Step 1, System Pattern Analysis, by representing the generative configurations that emerge from combinations of control frequency ($C$) and transparency level ($T$), thereby translating system-side reasoning into a reusable artifact for subsequent responsibility-gap analysis.
Control frequency includes two characteristic levels, high ($C_H$) and low ($C_L$), which describe how often human agents can intervene, adjust, or override generative processes. 
Transparency level likewise includes high ($T_H$) and low ($T_L$) conditions, representing the extent to which the model’s reasoning, uncertainty, and decision logic are interpretable to human observers. 
Each cell in the table corresponds to a distinct pattern of generative behavior that results from the interaction between autonomy and interpretability, and together these configurations define the basic responsibility space of GenAI-enabled systems.

\subsubsection{Human-Side Backbone Table ($A \times I$)}

The Human-Side Backbone Table \ref{tab:ai_backbone} serves as the documentation artifact for Step 2, Human Role Analysis, by structuring human oversight roles according to authority level ($A$) and interaction mode ($I$).
Authority includes three levels, operational ($A_1$), supervisory ($A_2$), and audit ($A_3$), representing the scope of decision power and accountability assigned to human agents. 
Interaction includes four types, active control ($I_1$), approval or validation ($I_2$), monitoring or auditing ($I_3$), and corrective or maintenance actions ($I_4$), describing how and when humans engage with system functions.
Together these dimensions describe how responsibility and oversight are structured across roles with different mandates and interaction modes.
Each cell in the table corresponds to a distinct role archetype that combines a particular authority level and interaction type.

Meanwhile, the table captures both the practical and normative constraints that shape valid role archetypes, distinguishing configurations that occur in practice from those that are theoretically possible but institutionally rare.
These configurations form a coherent typology of human oversight, in which five common roles emerge across contexts, including the Operator, the Maintainer, the Reviewer, the Coordinator, and the Auditor.
This typology provides the analytical foundation for mapping human involvement to system patterns in the alignment stage of the pipeline.

\begin{table*}[htbp]
\centering
\caption{Responsibility Gap Backbone Table ($P \times R$)}
\label{tab:pr_backbone}
\renewcommand{\arraystretch}{1.25}
\begin{tabular}{|P{2cm}|P{2.8cm}|P{2.8cm}|P{2.8cm}|P{2.8cm}|P{2.8cm}|}
\hline
\textbf{$P\times R$} & \textbf{Operator ($R_1$)} & \textbf{Reviewer ($R_2$)} & \textbf{Coordinator ($R_3$)} & \textbf{Maintainer ($R_4$)} & \textbf{Auditor ($R_5$)} \\ \hline
\textbf{Co-generation ($P_1$)} 
& \textbf{Procedural gap} — uncertainty about when to intervene or hand over control during shared generation cycles. 
& \textbf{Redundancy gap} — overlapping validation steps cause duplicated effort and unclear division of accountability. 
& \textbf{Coordination gap} — boundaries between operational and supervisory authority are undefined, leading to unclear oversight. 
& \textbf{Temporal gap} — maintenance updates lag behind active generation, creating misaligned responsibility over time. 
& \textbf{Trace gap} — frequent revisions obscure provenance and weaken auditability of collaborative outputs. \\ \hline
\textbf{Blind steering ($P_2$)} 
& \textbf{Epistemic gap} — operator makes interventions without understanding the model’s reasoning or underlying logic. 
& \textbf{Informational gap} — reviewer must approve opaque results without access to sufficient explanatory information. 
& \textbf{Delegation gap} — coordinator lacks visibility into model behavior, limiting capacity to allocate responsibility. 
& \textbf{Visibility gap} — maintainer corrects issues without causal insight into how or why they occurred. 
& \textbf{Traceability gap} — audit logs are incomplete, leaving the source and accountability chain uncertain. \\ \hline
\textbf{Review and approve ($P_3$)} 
& \textbf{Normative gap} — operator is accountable for outcomes but lacks the authority to influence the system’s behavior. 
& \textbf{Credibility gap} — reviewer depends on partial or unreliable model evidence when approving outputs. 
& \textbf{Escalation gap} — overlapping supervision tiers create confusion about who must respond to identified issues. 
& \textbf{Feedback gap} — delayed return of information blurs accountability for timely corrective action. 
& \textbf{Control gap} — auditor detects noncompliance but lacks the mandate or tools to intervene directly. \\ \hline
\textbf{Autonomous generation ($P_4$)} 
& \textbf{Disempowerment gap} — human remains accountable for outcomes produced by unseen or autonomous processes. 
& \textbf{Accountability gap} — reviewer formally approves automation results they have no authority to modify. 
& \textbf{Ownership gap} — distributed actors share responsibility, yet none assume clear accountability for outcomes. 
& \textbf{Reactive gap} — maintainer acts only after harm occurs, leaving preventive responsibility unassigned. 
& \textbf{Moral gap} — accountability is deferred to post hoc justification once consequences become visible. \\ \hline
\end{tabular}
\end{table*}

\subsubsection{Responsibility Gap Backbone Table ($P \times R$)}

The Responsibility Gap Backbone Table \ref{tab:pr_backbone} serves as the bridge between Step 3, Responsibility Gap Analysis, and Step 4, Requirements Derivation, by organizing each system pattern ($P$) and human role ($R$) pairing as a potential responsibility gap, from which corresponding human oversight requirements can be derived.
For each selected cell in the table, the identified responsibility gap is treated as an explicit input to the requirements-derivation step, where it is translated into one or more system-side or human-side requirements.
The resulting requirements retain traceability to the originating $P \times R$ pairing, so that each requirement can be traced back to the system configuration, human role, and responsibility gap that motivated it.
It integrates the system-side dimensions of control and transparency with the human-side dimensions of authority and interaction, allowing mismatches between autonomy, interpretability, and oversight to be systematically identified. 

The table is organized by four system patterns representing characteristic generative configurations, including co-generation ($P_1$: $C_H \times T_H$), blind steering ($P_2$: $C_H \times T_L$), review and approve ($P_3$: $C_L \times T_H$), and autonomous generation ($P_4$: $C_L \times T_L$). 
Across these, five common human roles are considered, including Operator ($R_1$), Reviewer ($R_2$), Coordinator ($R_3$), Maintainer ($R_4$), and Auditor ($R_5$). 
Each combination expresses a distinct form of potential responsibility gap, ranging from procedural ambiguities to normative or epistemic misalignments.
\section{User Study}
\label{sec:user_study}

The following research questions (RQs) guide the evaluation of the proposed approach across conceptual, methodological, and artifact dimensions identified in earlier sections.

\begin{itemize}
    \item \textbf{RQ1 (Conceptual):} To what extent does the proposed conceptual structure achieve completeness and consistency in representing human oversight within GenAI-enabled software?
    \item \textbf{RQ2 (Methodological):} To what extent does the Backbone-Anchored Deductive Pipeline ensure process traceability and feasibility in analyzing human oversight requirements?
    \item \textbf{RQ3 (Artifact):} To what extent does the Deductive Backbone Table provide representational traceability and conformance in documenting the links between conceptual dimensions, responsibility gaps, and derived oversight requirements?
\end{itemize}

\subsection{User Study Design}
\subsubsection{Design Overview}
The user study was designed as a controlled within-subject experiment to evaluate how the proposed methodology performs across its conceptual, methodological, and artifact layers.
Each participant completed two short requirements analysis tasks, applying two different approaches to two independent scenarios, enabling direct comparison under controlled conditions.
Accordingly, the evaluation combined two measures: participant questionnaire ratings to capture perceived improvement and expert ratings to evaluate the quality of the generated requirement artifacts.

\subsubsection{Variables and Controlled Factors}
\label{sec:variables}

\textbf{Independent variables.} 
The manipulated factor in this study is the analysis approach applied by participants. 
Two conditions were defined:
\begin{itemize}
    \item \textbf{Approach 1 (baseline GORE):} 
    A conventional goal-oriented requirements engineering (GORE) process following the five reasoning steps: identify goals, refine goals, assign agents, find obstacles, and derive requirements.
    \item \textbf{Approach 2 (enhanced GORE):} 
    The same five-step reasoning process enhanced by the Backbone-Anchored Deductive Pipeline, 
    which introduces explicit analysis of system patterns, human roles, and responsibility gaps prior to requirement derivation.
\end{itemize}

\textbf{Dependent variables.}
The dependent variables were organized into two groups corresponding to the two evaluation measures described above.

First, participant-perceived improvement was measured through the post-task questionnaire.
Each dimension captures a specific aspect of how the analysis approach influenced participant performance and perceived quality, and together they align with the three analytical layers and RQs of the study.
Participants evaluated each dimension (Completeness (RQ1), Consistency (RQ1), Process Traceability (RQ2), Feasibility (RQ2), Representational Traceability (RQ3), Conformance (RQ3), and Overall Effectiveness) through comparative statements of the form: "Approach 2 is better than Approach 1 in terms of~[dimension]."
Responses were recorded on a five-point Likert scale 
(1 = Strongly Disagree, 5 = Strongly Agree), 
where higher scores indicate stronger perceived improvement of the proposed approach.

Second, expert-rated artifact quality was measured through independent expert assessment of the generated requirement artifacts. 
To maintain consistency with the evaluation framework, the expert rating reused the same six evaluation attributes: Completeness, Consistency, Process Traceability, Feasibility, Representational Traceability, and Conformance.
Unlike the participant questionnaire, which measured perceived improvement between approaches, the expert rating applied these attributes directly to the generated requirement artifacts.
The assessment was conducted by two independent experts who did not participate in the user study, with the detailed expert-rating procedure described in Section \ref{sec:collection}.

\textbf{Controlled factors.}
The following factors were managed to maintain internal validity rather than treated as variables.  
Specifically, two GenAI-enabled clinical documentation scenarios were used as comparable task contexts within the medical domain. 
Additionally, task order was counter-balanced across participants to mitigate potential learning effects, and participant profiles were kept homogeneous in background and expertise to ensure that observed differences in results could be attributed primarily to the analysis approach.

\subsubsection{Participants and Materials}

\textbf{Participants.}
A total of 24 participants took part in the user study.
They were recruited from academic and professional communities across Hong Kong (7), Mainland China (5), Japan (9), and Australia (3), providing an internationally diverse sample of analysts and practitioners.
Participants included computer science students (17), software engineers (3), quantitative analysts (2), and early-career researchers or scholars (2) in computing-related fields.
This composition ensured that all participants possessed basic analytical and problem-solving competence relevant to requirements reasoning tasks, while differing in domain specialization and professional context.
Participation was voluntary and uncompensated, and all responses were anonymized. 

\textbf{Materials.}  
Each participant completed two analytical tasks based on GenAI-enabled scenarios designed to vary the pattern of GenAI--human interaction and human oversight required, including Scenario A, Interactive Treatment-Plan Assistant, and Scenario B, Automated Follow-up Summary Generator.
Detailed scenario descriptions are provided in Appendix~\ref{appendix:scenarios}.

\subsubsection{Design and Procedure}
\label{sec:procedure}

\textbf{Design.}  
Participants were randomly assigned to two balanced groups. 
Group~A completed Scenario~A using the GORE approach and Scenario~B using the enhanced GORE approach, whereas Group~B performed the same two scenarios in the reverse order. 
This design enabled direct comparison between approaches while maintaining equivalence in scenario exposure and task complexity.

\textbf{Procedure.}  
Each participant completed the entire study individually using an online form that presented the two scenarios sequentially. 
The session followed the sequence below:

\begin{enumerate}
    \item \textit{Briefing and consent.} Participants reviewed the study information, provided informed consent, and supplied basic background details (e.g., demographics, experience, and familiarity with SE/RE and GenAI).
    \item \textit{Task 1.} Participants analyzed the assigned scenario using the standard GORE approach. 
    Instructions for the five reasoning steps were provided inline within the online form to guide the analysis.
    \item \textit{Task~2.} Participants analyzed the remaining scenario using the enhanced GORE approach. They first conducted the proposed Backbone-Anchored Deductive Pipeline analysis by identifying the system pattern, human roles, and potential responsibility gaps, and then proceeded through the five goal-oriented reasoning steps.
    \item \textit{Comparative questionnaire.} After both tasks, participants completed a short comparative questionnaire to evaluate the two approaches based on the dimensions introduced in Section~\ref{sec:variables}. 
\end{enumerate}

The two tasks were conducted under identical conditions across participants, with materials, instructions, and task order controlled as described earlier.
Each session lasted approximately 40–60 minutes.

\subsubsection{Data Collection and Validity Considerations}
\label{sec:collection}

\textbf{Participant-perception data.}
Questionnaire and time data were collected from the post-task comparative questionnaire and the self-reported completion times recorded in the online form.
All questionnaire responses were recorded using the same five-point Likert scale with anchored descriptions introduced in Section \ref{sec:variables}.
Responses on the six evaluation dimensions plus one overall effectiveness item were summarized using descriptive statistics (mean and standard deviation (SD)) to characterize central tendencies and variation across participants.
The additional item measuring overall effectiveness is used to capture participants’ general perception of improvement when applying the enhanced GORE approach.
For interpretive clarity, a mean score above 3.0 indicated a generally favorable perception of the enhanced approach relative to the baseline.
Self-reported task times were analyzed descriptively to provide complementary evidence regarding procedural feasibility.
To strengthen conclusion validity, participant-perception ratings were further analyzed using one-sample Wilcoxon signed-rank tests against the neutral midpoint of 3. 
This test was selected because the questionnaire data were ordinal and the comparative statements measured whether participants perceived the enhanced GORE approach as better than the baseline. 
Effect sizes were reported using rank-biserial correlation. 
The internal consistency of the six quality attribute dimensions was assessed using Cronbach's alpha.

\textbf{Expert-rating data.} Expert-rating data were collected from the independent assessment of the generated requirement artifacts. 
The assessment was conducted by two independent experts who did not participate in the user study and were blind to the study conditions.
It proceeded in two sequential stages.
First, both experts independently rated the complete set of generated requirement artifacts using the six evaluation attributes and the additional overall artifact-quality item on a five-point ordinal scale, without consulting each other during this stage.
After both independent assessments were completed, the experts compared their ratings and reviewed cases in which their scores differed.
They discussed the rationale for these discrepant ratings until consensus scores were reached, with higher scores indicating higher artifact quality on the corresponding attribute.
The resulting consensus scores were used in the subsequent expert-rating analysis.
The expert-rating results were summarized descriptively by comparing the mean ratings of artifacts produced using the baseline and enhanced approaches.

\textbf{Qualitative data.}  
The questionnaire included an optional open-ended prompt inviting participants to comment on the strengths or limitations of the two approaches.
These comments were reviewed through light content analysis to identify recurring perceptions related to clarity, traceability, or effort.
Given the small and uneven number of responses, the qualitative findings were used solely to contextualize the quantitative results rather than as independent evaluation metrics.

\textbf{Validity considerations.}  
Construct validity was supported by mapping each evaluation dimension directly to the corresponding RQ and by triangulating participant perceptions with artifact-level evidence.
Internal validity was maintained through the within-subject, counter-balanced design and consistent materials across conditions.
Because each participant applied the two approaches to different scenarios, scenario-specific differences may still influence artifact-level comparisons.
To mitigate this threat, the two scenarios were designed to be comparable in domain, size, and task complexity, and the counter-balanced assignment ensured that both scenarios were represented under both approaches across participants.
External validity benefited from the internationally diverse yet analytically homogeneous participant group.
Conclusion validity was strengthened by incorporating artifact-level evidence, reliability assessment, and inferential comparison, while claims were still interpreted cautiously given the limited sample size and controlled task setting.
All anonymized data and supplementary materials are available in the project’s public repository.

\subsection{Results and Analysis}

\begin{figure*}[t]
\centering

\pgfplotstableread[row sep=\\,col sep=&]{
Dimension & SD & D & N & A & SA \\
Completeness & 0 & 0 & 4.2 & 58.3 & 37.5 \\
Consistency & 0 & 0 & 8.3 & 79.2 & 12.5 \\
Process Traceability & 4.2 & 8.3 & 4.2 & 41.7 & 41.7 \\
Feasibility & 0 & 4.2 & 41.7 & 41.7 & 12.5 \\
Representational Traceability & 0 & 0 & 8.3 & 37.5 & 54.2 \\
Conformance & 0 & 0 & 12.5 & 58.3 & 29.2 \\
Overall Effectiveness & 0 & 0 & 12.5 & 41.7 & 45.8 \\
}\likertdata

\begin{tikzpicture}
\begin{axis}[
    xbar stacked,
    width=0.75\linewidth,
    height=12cm,
    xmin=0, xmax=100,
    xlabel={Percentage},
    symbolic y coords={
        Completeness,
        Consistency,
        Process Traceability,
        Feasibility,
        Representational Traceability,
        Conformance,
        Overall Effectiveness},
    ytick=data,
    y dir=reverse,
    xtick={0,20,40,60,80,100},
    bar width=7pt,
    enlarge y limits=0.1,
    tick label style={font=\small},
    label style={font=\small},
    legend style={
        font=\footnotesize,
        legend columns=5,
        at={(0.5,-0.18)},
        anchor=north,
        draw=none
    },
    legend cell align=left,
    legend style={column sep=6pt},
]

\definecolor{c1}{HTML}{C75A23}
\definecolor{c2}{HTML}{EDB184}
\definecolor{c3}{HTML}{C9D8A3}
\definecolor{c4}{HTML}{6D8B3F}

\addplot+[fill=c1] table[x=SD, y=Dimension] {\likertdata};
\addplot+[fill=c2] table[x=D,  y=Dimension] {\likertdata};
\addplot+[fill=gray!10] table[x=N,  y=Dimension] {\likertdata};
\addplot+[fill=c3] table[x=A,  y=Dimension] {\likertdata};
\addplot+[fill=c4] table[x=SA, y=Dimension] {\likertdata};

\legend{1 Strongly Disagree, 2 Disagree, 3 Neutral, 4 Agree, 5 Strongly Agree}

\end{axis}
\end{tikzpicture}

\caption{Distribution of participant responses across evaluation dimensions (five-point Likert scale; statement: “Approach 2 is better than Approach 1 in terms of [dimension]. ”)}
\label{fig:likert}
\end{figure*}
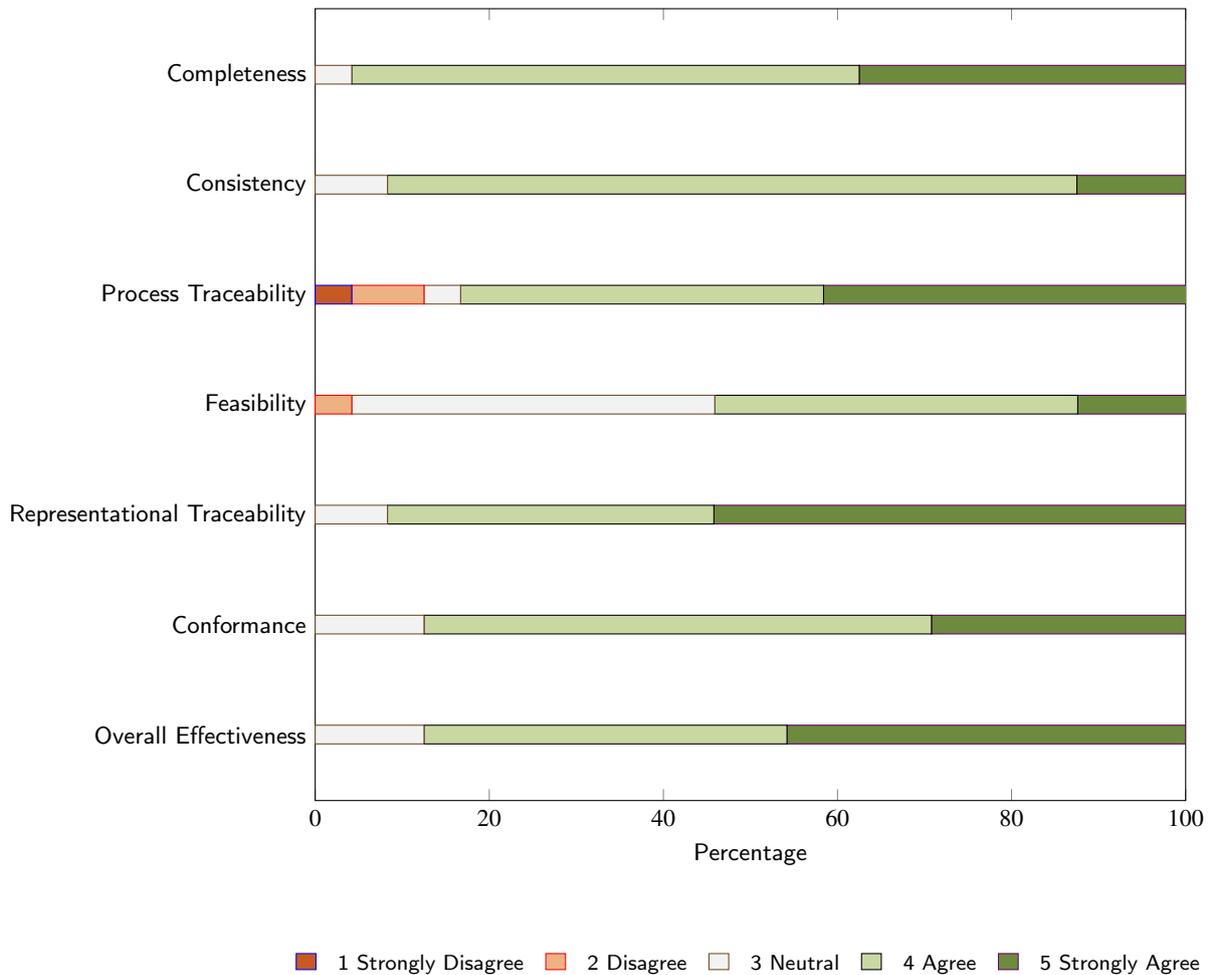

This subsection presents the results of the controlled user study and analyzes them with respect to the three RQs. 
Each RQ corresponds to one analytical layer of the proposed methodology, conceptual (RQ1), methodological (RQ2), and artifact-level (RQ3), and is evaluated through two associated dimensions derived from the requirement quality attributes defined in the relevant standard. 
The results are organized into three parts: participant demographics, results for RQ1–RQ3, and an integrated discussion of key findings.
Figure~\ref{fig:likert} presents the distribution of participant-perception data across all evaluation dimensions using a 100\% stacked bar chart. 
The overall effectiveness item received broadly positive ratings (mean = 4.33, SD = 0.70), indicating that participants viewed the enhanced GORE approach as generally more effective across the evaluated dimensions.
The six evaluation dimensions showed acceptable internal consistency (Cronbach's \(\alpha = 0.736\)). Therefore, the dimensions were interpreted individually according to their corresponding RQs rather than combined into a single composite score. 
The one-sample Wilcoxon signed-rank tests further showed that all participant-perception ratings were significantly above the neutral midpoint of 3, with large rank-biserial effect sizes (Table~\ref{tab:wilcoxon-participant}).
In addition to participant-perception results, the analysis reports expert ratings of the generated requirement artifacts. 
As shown in Table~\ref{tab:expert-ratings}, artifacts produced using the enhanced GORE approach received higher mean expert ratings than baseline artifacts on six of the seven artifact-quality items, with the largest descriptive improvements observed for consistency and representational traceability.
These results should be interpreted as descriptive patterns in the experts’ assessments of the evaluated artifacts, rather than as conclusions extending beyond the artifacts examined in this study.

\subsubsection{Participant Demographics}

All 24 recruited participants completed both analytical tasks, yielding a complete set of valid responses.
Two-thirds of the participants identified as male (66.7\%) and one-third as female (33.3\%). 
Most participants were aged between 25-34 years (62.5\%), followed by those aged 18-24 years (33.3\%). 
In terms of education, most participants held a postgraduate degree, with 50\% having completed a master’s and 12.5\% a doctoral qualification, while 37.5\% held a bachelor’s degree.
Regarding professional experience, all participants reported prior exposure to software or RE concepts, and 20.8\% had less than one year of experience, 41.7\% had one to two years, 29.2\% had three to five years, and 8.3\% had more than five years. 
Self-reported familiarity with goal-oriented requirements engineering was generally moderate, with most participants selecting level 3 on a five-point scale. 
All participants indicated previous use of generative AI tools for coursework or research, while 54.2\% had developed or experimented with GenAI software and 41.7\% had conducted or supervised research involving GenAI systems. 
No missing or incomplete data were recorded, and no participant reported prior exposure to the proposed methodology.

\begin{table*}[t]
\centering
\caption{Expert ratings of generated requirement artifacts under the baseline and enhanced approaches. Values report mean (M), standard deviation (SD), and mean difference (Enhanced M -- Baseline M) on a five-point ordinal scale (1 = Very poor, 5 = Excellent).}
\label{tab:expert-ratings}
\begin{tabular}{lccccc}
\hline
\textbf{Attribute} & \textbf{Baseline M} & \textbf{Baseline SD} & \textbf{Enhanced M} & \textbf{Enhanced SD} & \textbf{Difference} \\
\hline
Completeness & 3.63 & 0.77 & 3.67 & 0.70 & +0.04 \\
Consistency & 3.67 & 0.82 & 4.46 & 0.78 & +0.79 \\
Process Traceability & 3.79 & 0.78 & 3.71 & 0.86 & -0.08 \\
Feasibility & 3.75 & 0.68 & 3.96 & 0.55 & +0.21 \\
Representational Traceability & 3.67 & 0.92 & 4.46 & 0.83 & +0.79 \\
Conformance & 4.21 & 0.78 & 4.58 & 0.58 & +0.38 \\
Overall Artifact Quality & 3.71 & 0.81 & 3.88 & 0.80 & +0.17 \\
\hline
\end{tabular}
\end{table*}

\subsubsection{Results for RQ1}

\textbf{Participant-perception results.} 
RQ1 examines the conceptual layer of the proposed methodology, focusing on the dimensions of completeness and consistency. 
As shown in Figure~\ref{fig:likert}, participants rated the enhanced approach (Approach 2) notably higher than the baseline goal-oriented process (Approach 1) on both dimensions. 
The mean rating for completeness was 4.33 (SD = 0.56), and for consistency 4.04 (SD = 0.46), reflecting strong agreement that the proposed methodology offers a more comprehensive and coherent conceptual foundation for oversight analysis. 
Most participants (over 90\%) selected "agree" or "strongly agree" on both dimensions, and none provided a rating below neutral.
These patterns indicate strong consensus that the enhanced methodology provides a more systematic conceptual foundation for reasoning about oversight requirements.
The detailed distribution of responses across all dimensions is provided in Table~\ref{tab:dimension_distribution}.

\textbf{Expert-rating data.} 
The expert-rating results provide artifact-level evidence for the two RQ1 attributes. 
As shown in Table~\ref{tab:expert-ratings}, artifacts produced using the enhanced GORE approach received a slightly higher mean rating for completeness than baseline artifacts (3.67 vs. 3.63), and a substantially higher mean rating for consistency (4.46 vs. 3.67).

\textbf{Qualitative observations.} 
No substantial critical feedback was raised regarding conceptual clarity or coverage. 
Participant submissions generally reflected coherent reasoning and appropriate differentiation of system and human oversight roles, indicating that the enhanced process was well understood and consistently applied. 

\textbf{Interpretation.}
These results suggest that the proposed methodology improves the conceptual quality of oversight representation, especially in terms of consistency. 
The participant-perception results indicate that analysts regarded the enhanced approach as more complete and coherent, while the expert-rating results provide artifact-level support primarily for improved consistency. 
This appears to result from the method's explicit structuring of key conceptual elements, such as where human intervention is possible, what forms of transparency are required, and which contextual dependencies matter. 
By externalizing these conceptual distinctions, the methodology helps analysts align system-side patterns, human-side roles, responsibility gaps, and derived requirements within a clearer analytical structure. 
However, the relatively small expert-rated difference in completeness suggests that the method's main benefit in this study was not simply increasing the number or breadth of identified oversight concerns, but improving the coherence with which those concerns were represented and connected.

\subsubsection{Results for RQ2}

\textbf{Participant-perception results.} 
RQ2 investigates the methodological layer of the proposed methodology, focusing on process traceability and feasibility. 
As shown in Figure~\ref{fig:likert}, participants rated the enhanced approach (Approach 2) higher than the baseline goal-oriented process (Approach 1) in both dimensions. 
The mean rating for process traceability was 4.08 (SD = 1.10), indicating agreement that the deductive structure improved transparency and linkage between analytical steps. 
The mean rating for feasibility was 3.63 (SD = 0.77), suggesting moderate agreement that the method was practically applicable within the study setting. 
Most participants (around 83\%) selected "agree" or "strongly agree" for process traceability, while over half (around 54\%) responded positively for feasibility. 
The detailed distribution of responses across all dimensions is provided in Table \ref{tab:dimension_distribution}.

In addition to subjective comments, the recorded task durations provide an objective indicator of feasibility. 
As shown in Table \ref{tab:time_comparison}, the mean completion time for Approach 1 (baseline) was 29.13~ minutes (SD = 8.15), while Approach 2 (enhanced) averaged 30.58 minutes (SD = 8.38), with a mean difference of only 1.46 minutes. 
The individual times in Table \ref{tab:completion_time} show that some participants completed the task more quickly using the baseline approach, whereas others completed it more quickly using the enhanced approach, with no consistent pattern across participants.
This distribution indicates that the enhanced process does not introduce a substantial time burden relative to the baseline.

\begin{table}[h!]
\centering
\caption{Comparison of task completion time (minutes)}
\label{tab:time_comparison}
\begin{tabular}{P{4cm}P{1.5cm}P{1.5cm}}
\hline
\textbf{Approach} & \textbf{Mean} & \textbf{SD}\\ \hline
Approach~1 (Baseline) & 29.13 & 8.15 \\
Approach~2 (Enhanced) & 30.58 & 8.38 \\ \hline
\end{tabular}
\end{table}

\begin{table}[t]
\centering
\caption{One-sample Wilcoxon signed-rank test results for participant-perception ratings against the neutral midpoint of 3. Effect sizes are reported using rank-biserial correlation.}
\label{tab:wilcoxon-participant}
\begin{tabular}{lcccc}
\hline
\textbf{Dimension} & \textbf{Median} & \(\mathbf{W^+}\) & \(\mathbf{p}\) & \(\mathbf{r_{rb}}\) \\
\hline
Completeness & 4 & 276 & $< .001$ & 1.00 \\
Consistency & 4 & 253 & $< .001$ & 1.00 \\
Process Traceability & 4 & 245 & $< .001$ & 0.78 \\
Feasibility & 4 & 99 & $< .001$ & 0.89 \\
Representational Traceability & 5 & 253 & $< .001$ & 1.00 \\
Conformance & 4 & 231 & $< .001$ & 1.00 \\
Overall Effectiveness & 4 & 231 & $< .001$ & 1.00 \\
\hline
\end{tabular}
\end{table}

\textbf{Expert-rating data.}
As shown in Table~\ref{tab:expert-ratings}, artifacts produced using the enhanced GORE approach received a slightly lower mean rating for process traceability than baseline artifacts (3.71 vs. 3.79), but a higher mean rating for feasibility (3.96 vs. 3.75). 
This suggests that the artifact-level evidence for RQ2 is stronger for feasibility than for process traceability.

\textbf{Qualitative observations.} 
Participant feedback predominantly centered on the understandability, usability, and perceived time efficiency of the enhanced process.
Participants acknowledged that Approach 2 provided a clearer and more traceable reasoning structure, however, they also highlighted that the structured process initially imposed a steeper learning curve. 
Representative comments include, "the current learning curve is challenging," "too long time required for approach 2," and "providing a quick reference cheat-sheet for each step."
Two participants further suggested adding brief examples or simplified explanations for technical terms to enhance readability. 
Overall, these comments indicate that the perceived complexity was primarily cognitive rather than procedural.

\textbf{Interpretation.}
Participants perceived the enhanced approach as improving the transparency and linkage of the reasoning process, and the comparable task completion times suggest that this added structure did not introduce a substantial procedural burden.
The expert-rating results, however, indicate that this perceived improvement in process traceability was not consistently reflected in the final generated artifacts, as expert-rated process traceability remained slightly lower than the baseline.
This divergence suggests that the enhanced approach may have supported participants' reasoning during task execution, but that the resulting reasoning chains were not always fully preserved or made explicit in the submitted artifacts. 
In contrast, the higher expert-rated feasibility score indicates that the enhanced approach helped produce requirements that were judged as somewhat more practically applicable within the scenario context. 
Overall, the results suggest that the Backbone-Anchored Deductive Pipeline improved methodological guidance and feasibility, while further refinement is needed to ensure that the intermediate reasoning process is more consistently documented in the final artifacts.

\subsubsection{Results for RQ3}

\textbf{Participant-perception results.} 
RQ3 examines the artifact layer of the proposed methodology, evaluating representational traceability and conformance in documenting the reasoning process and derived oversight requirements. 
As shown in Figure~\ref{fig:likert}, participants rated the enhanced approach (Approach 2) higher than the baseline goal-oriented process (Approach 1) on both dimensions. 
The mean rating for representational traceability was 4.46 (SD = 0.66), reflecting strong agreement that the Deductive Backbone Table and associated documentation improved the visibility of reasoning links between conceptual elements, responsibility gaps, and derived requirements. 
The mean rating for conformance was 4.17 (SD = 0.64), indicating that participants found the artifact structure consistent with the intended oversight representation format. 
Most participants (over 90\%) selected "agree" or "strongly agree" for representational traceability, and 87.5\% provided positive ratings for conformance. 
The detailed distribution of responses across all dimensions is provided in Table \ref{tab:dimension_distribution}.

\textbf{Expert-rating results.}
As shown in Table~\ref{tab:expert-ratings}, artifacts produced using the enhanced GORE approach received substantially higher mean ratings for representational traceability than baseline artifacts (4.46 vs. 3.67), and higher mean ratings for conformance (4.58 vs. 4.21). 
This indicates that the strongest artifact-level improvement was observed in representational traceability, while conformance also improved despite the relatively high baseline score.

\textbf{Qualitative observations.} 
Participant feedback on artifact quality primarily concerned the interpretability of the documentation.
One participant mentioned that "for the explanation of the meaning of professional terms, a simple example can be given," indicating that while the artifact’s structure was effective, some of its terminology could be made more accessible.
This comment suggests that interpretability could be improved through the inclusion of illustrative examples or concise definitions, particularly for readers less familiar with RE formalisms.

\textbf{Interpretation.} 
Participants perceived the enhanced approach as improving both representational traceability and conformance, and the expert-rating results show corresponding artifact-level improvements on both attributes. 
The particularly strong improvement in representational traceability suggests that the Deductive Backbone Table helped preserve explicit links among conceptual elements, responsibility gaps, and derived oversight requirements. 
The improvement in conformance further indicates that the enhanced artifacts were judged as more closely aligned with the intended documentation structure of the proposed methodology. 
These findings support the role of the Deductive Backbone Table as a reusable representation for documenting responsibility-gap analysis and oversight requirement derivation. 
At the same time, the qualitative feedback suggests that the artifact's interpretability could be further improved by adding concise definitions or examples for technical terms, especially for users less familiar with RE formalisms.

\subsubsection{Discussion and Limitations}

\textbf{Divergence in process traceability results.}
A notable exception to the generally positive pattern concerns process traceability. 
Although participants generally perceived the enhanced GORE approach as improving the transparency and linkage of the reasoning process, several participants still rated the enhanced approach below neutral on this dimension. 
In addition, the expert-rating results showed a slight decrease in process traceability for the enhanced artifacts compared with the baseline artifacts.
This divergence may be explained by three related factors. 
First, the Backbone-Anchored Deductive Pipeline is not fully linear: system pattern analysis and human role analysis are conducted as two parallel analytical inputs before being integrated during responsibility-gap analysis. 
Although this structure reflects the conceptual logic of the methodology, it may be more difficult for participants to document as a single sequential reasoning chain. 
Second, the enhanced approach introduces more intermediate reasoning constructs. 
This added analytical depth may have improved conceptual and artifact-level quality, but it also increased the burden of preserving explicit links among all intermediate steps in the final written artifacts. 
Third, the study instructions may not have sufficiently emphasized how participants should record trace links across the parallel and sequential parts of the pipeline. 
As a result, participants may have used the pipeline to guide their reasoning internally while leaving some intermediate connections implicit in their submitted artifacts. 
A priority for future work is therefore to compare alternative instruction formats and artifact templates to determine whether the observed process traceability divergence reflects a limitation of the methodology itself or instead resulted from the user-study materials.

\textbf{Simplified representation of control and transparency.}
As acknowledged in Section~\ref{sec:methodology}, the methodology intentionally models each dimension as a single analytical attribute within the deductive backbone to maintain a concise and tractable structure.
Participant reasoning artifacts and feedback revealed that perceptions of transparency and control varied across analytical layers: some participants associated transparency with clearer traceability of reasoning, whereas others referred to the interpretability of generated outputs or human decision visibility. 
Similarly, control was largely treated as a binary capacity for human intervention rather than as a continuum spanning procedural oversight, adaptive adjustment, and real-time response.
This abstraction proved sufficient for evaluating conceptual completeness and consistency (RQ1) but constrains applicability in complex oversight settings where these layers interact dynamically.
Future work will refine the treatment of control and transparency to incorporate such sub-dimensions while preserving the backbone’s analytical coherence and methodological clarity.

\textbf{Dependence on subjective expert judgments.}  
The characterization of responsibility gaps and the assessment of oversight attributes relied on participants’ qualitative interpretations of design information and role descriptions. 
Variation in feasibility ratings (SD = 0.77, the second highest among all dimensions) and several qualitative comments on learning effort and clarity suggest interpretive differences among analysts. 
Such reliance on individual judgment, while inherent to requirements reasoning, introduces subjectivity that may affect methodological consistency (RQ2). 
To mitigate this, future work should incorporate more objective reference schemes, such as standardized audit checklists or benchmark cases, and include multiple stakeholder perspectives (e.g., developers, auditors, and end-users) to establish shared criteria for evaluating control and transparency.

\textbf{Organizational and social boundary conditions.}  
The methodology focuses on analytical traceability and representational conformance but does not directly address how oversight responsibilities are enacted within organizational or social contexts. 
The controlled study deliberately abstracted away contextual influences such as managerial authority, cultural norms, or accountability distribution. 
However, prior studies and practitioner reports indicate that organizational structures can reshape or even displace accountability, for example, transferring liability for AI errors to designated human reviewers despite transparent documentation \cite{johnson2019ai}.
This limitation reflects the boundary of analytical evaluation rather than a deficiency in methodology design.
Future research should therefore investigate how the documented responsibility structures interact with governance mechanisms, incentive systems, and institutional culture to ensure that representational accountability translates into equitable and enforceable practice.

\textbf{Limited industrial and cross-domain validation.}
The present study evaluates the proposed methodology through controlled analytical tasks rather than larger-scale validation in real industrial projects.
This design supports focused comparison between the baseline and enhanced approaches, but it limits the extent to which the findings can demonstrate practical effectiveness in organizational settings.
In addition, the experimental scenarios were limited to selected GenAI-enabled application contexts, which may not fully represent the diversity of domains in which responsibility gaps arise.
In real GenAI-enabled software projects, responsibility-gap analysis may be influenced by additional factors, such as stakeholder negotiation, time pressure, regulatory constraints, toolchain integration, domain-specific risks, and organizational accountability structures.
Therefore, the results should be interpreted as initial evidence of the methodology’s analytical usefulness and perceived value, rather than as conclusive evidence of industrial or cross-domain effectiveness.
Future work should evaluate the methodology in larger and more realistic project settings, including real GenAI-enabled software development or governance contexts across multiple domains.

\subsection{Threats to Validity}

This subsection outlines potential factors that may threaten the validity of the study across four dimensions.

\subsubsection{Construct Validity}
Construct validity may be threatened if participants interpreted the evaluation dimensions differently from the intended meaning. 
Although the measures were derived from ISO/IEC/IEEE~29148:2018 requirement quality attributes, participants' subjective understanding of concepts such as "completeness", "traceability," or "conformance" could still vary. 
For each attribute, a brief explanatory sentence was provided in the questionnaire to clarify its intended meaning, but individual differences in interpretation may nevertheless occur. 
Such variation could lead to inconsistent use of the Likert scale, making the scores reflect perception rather than the intended construct of oversight quality.

\subsubsection{Internal Validity}
Internal validity could be affected by residual differences between the two task scenarios used in the within-subject design. 
Because each participant applied the two approaches to distinct scenarios, any observed performance difference may partly stem from scenario-specific complexity rather than methodological effect. 
Learning or fatigue effects might also influence results, as participants could perform differently on the second task after becoming more familiar with the reasoning process.

\subsubsection{External Validity}
External validity may be limited by the composition of the participant group, domain coverage, and the experimental setting.
Although participants were recruited from multiple regions and all had prior exposure to software or RE concepts, the sample was still dominated by students and early-career participants, with relatively fewer industrial practitioners.
This limits the extent to which the findings can be generalized to experienced requirements engineers, project managers, compliance officers, or domain experts working on deployed GenAI-enabled systems.
In addition, the study used controlled textual scenarios within a limited set of GenAI-enabled application contexts rather than real project artifacts from diverse domains.
While these scenarios helped ensure comparability across participants, they cannot fully capture the complexity of industrial RE practice, such as stakeholder negotiation, evolving requirements, tool support, organizational constraints, and domain-specific accountability mechanisms.

\subsubsection{Conclusion Validity}
Conclusion validity may be threatened by the small sample size and the limited statistical power of the analysis. 
To mitigate this threat, we supplemented descriptive statistics with one-sample Wilcoxon signed-rank tests, effect-size estimates, and reliability assessment of the participant-perception questionnaire. 
Nevertheless, statistical generalization remains limited, and the results should be interpreted as initial evidence rather than conclusive proof of effectiveness.
In addition, the interpretation of open-ended comments during qualitative coding may introduce researcher subjectivity, potentially affecting the consistency of derived themes. 
The open-ended question in the questionnaire was optional, meaning that not all participants provided comments, which may lead to self-selection bias and limit the representativeness of qualitative insights.
\section{Conclusion}
\label{sec:conclusion}

This study proposes a design methodology for understanding, identifying, and documenting responsibility gaps in GenAI-enabled software from a human oversight requirements perspective.
By integrating responsibility gap analysis into established RE practices, the methodology enables the systematic conceptualization, derivation, and documentation of oversight requirements grounded in the distribution of responsibility between human and generative components.
The multi-layer methodology, comprising conceptualization, methodological, and artifact levels, supports a consistent process for defining responsibility elements, analyzing their interactions to reveal potential gaps and derive oversight requirements, and representing the resulting reasoning trace in a structured and reusable form.
In a controlled user study comparing the proposed methodology with a baseline goal-oriented RE process, participant-perception results indicated generally favorable evaluations of the enhanced approach across the six RQ-aligned dimensions.
Expert-rating results further provided artifact-level evidence that the enhanced approach improved several quality characteristics of the generated requirement artifacts, especially consistency and representational traceability.

Nevertheless, these findings should be interpreted in light of the student- and early-career-dominated sample and the use of controlled scenarios.
The methodology has not yet been validated in industrial GenAI-enabled software projects, limiting the generalizability of present evidence.

Future work will focus on refining the methodology and validating it in broader, more realistic settings.
First, the internal structure of the Backbone-Anchored Deductive Pipeline should be improved to make process traceability more explicit in the final artifacts.
Second, the representation of control and transparency should be refined to capture more fine-grained sub-dimensions, such as procedural oversight, real-time intervention, evidential transparency, and interpretability of generated outputs, while preserving the backbone’s analytical clarity.
Third, future studies should involve larger and more diverse practitioner groups and multiple stakeholder roles, such as requirements engineers, developers, auditors, domain experts, and end-users, to establish more objective assessment criteria and reduce reliance on individual interpretation.
Finally, the methodology should be evaluated in real GenAI-enabled software development or governance contexts across multiple domains to examine whether the documented responsibility structures remain useful under practical constraints.

\section*{Declaration of competing interest}
The authors declare that they have no known competing financial interests or personal relationships that could have appeared to influence the work reported in this paper.

\section*{Declaration of generative AI and AI-assisted technologies in the manuscript preparation process}
During the preparation of this work the author(s) used generative AI in order to refine language and improve clarity. After using this tool/service, the author(s) reviewed and edited the content as needed and take(s) full responsibility for the content of the published article.

\section*{Acknowledgment}
This work was partially conducted during the author's visit to Waseda University.

\section*{Data availability}
All anonymized user study responses and related materials are publicly available at \url{https://github.com/rockmao45/HOR}.

\printcredits

\bibliographystyle{plain}

\bibliography{cas-refs}

\clearpage
\appendix

\section{Scenario Descriptions}
\label{appendix:scenarios}

This section includes the two GenAI-enabled healthcare scenarios used in the user study described in Section~\ref{sec:user_study}. 

\subsection*{Scenario~A: Interactive Treatment-Plan Assistant}
A hospital uses a GenAI-based assistant to help physicians create personalized treatment plans for patients with chronic conditions.
During each consultation, the attending physician enters the patient’s current data and preferences into the system. 
The AI assistant immediately proposes several treatment options and shows its reasoning, confidence levels, and relevant medical-guideline references on an explanation panel.
The physician can adjust the inputs, change assumptions, or ask the assistant for alternative suggestions.
The system updates its recommendations right away and shows how each change affects expected outcomes.
The physician and the assistant work together until a suitable plan is reached, and the physician formally approves and records the final treatment plan.

\subsection*{Scenario~B: Automated Follow-up Summary Generator}
A hospital has introduced a GenAI-based system to support patient communication after discharge.
After each hospitalization, the AI system gathers information from patient records, discharge notes, and recent progress reports.
It automatically drafts follow-up summaries that include medication instructions, recovery reminders, and contact information for further assistance.
The messages are delivered to patients through the hospital information system as part of the routine follow-up process.
The attending physician and nursing staff can access general follow-up statistics and occasionally review selected summaries for quality assurance.
Routine message generation and delivery take place automatically without requiring direct human input for each case.
The clinical oversight coordinator monitors overall performance through periodic reports that include message completion rates and patient feedback indicators.
The system operates largely on its own to manage patient follow-up communication, while clinical staff provide oversight through scheduled reviews and quality monitoring.

\section{Participant Demographics and Backgrounds}

Table~\ref{tab:demographics} summarizes the demographic and background information of 24 participants who completed the user study. 
Percentages are calculated over all valid responses (\(n=24\)). 
All data were self-reported and fully anonymized.

\begin{table}[h]
\centering
\caption{Summary of participant demographics and backgrounds}
\label{tab:demographics}
\begin{tabular}{|P{2cm}|P{3.8cm}|P{1.5cm}|P{1cm}|}
\hline
\textbf{Category} & \textbf{Response Option} & \textbf{Percentage (\%)} & \textbf{Count} \\ \hline

Gender & Male & 66.7 & 16 \\
 & Female & 33.3 & 8 \\ \hline

 & 18-24 & 33.3 & 8 \\
Age Range & 25-34 & 62.5 & 15 \\
 & 35-44 & 4.2 & 1 \\ \hline

 & Bachelor’s degree or equivalent & 37.5 & 9 \\
Highest Education & Master’s degree or equivalent & 50.0 & 12 \\
 & Doctoral degree (Ph.D. or equivalent) & 12.5 & 3 \\ \hline

Years of & Less than 1 year & 20.8 & 5 \\
Experience in & 1–2 years & 41.7 & 10 \\
SE or RE & 3–5 years & 29.2 & 7 \\
Concepts & More than 5 years & 8.3 & 2 \\ \hline

 & Level 1 (Not familiar) & 20.8 & 5 \\
Familiarity & Level 2 & 25.0 & 6 \\
with Goal- & Level 3 & 41.7 & 10 \\
Oriented RE & Level 4 & 12.5 & 3 \\
 & Level 5 (Very familiar) & 0.0 & 0 \\ \hline

 & Used GenAI tools for coursework or research & 100.0 & 24 \\
Experience with GenAI & Developed or experimented with GenAI software & 54.2 & 13 \\
 & Conducted or supervised research in GenAI systems & 41.7 & 10 \\ \hline

\end{tabular}
\end{table}

\section{Distribution of Participant Ratings}

Table~\ref{tab:dimension_distribution} presents the distribution of participant ratings 
on the 1-5 Likert scale across the six evaluation dimensions and the overall metric used 
to assess the two approaches., 
along with the corresponding mean (M) and standard deviation (SD) for each dimension. 
Percentages are calculated over all valid responses (\(n=24\)). 

\begin{table}[h!]
\centering
\caption{Distribution of participant ratings}
\label{tab:dimension_distribution}
\begin{tabular}{|P{2.5cm}|P{0.5cm}|P{0.5cm}|P{0.5cm}|P{0.5cm}|P{0.5cm}|P{1cm}|P{1cm}|}
\hline
\textbf{Evaluation Dimension} & \textbf{1} & \textbf{2} & \textbf{3} & \textbf{4} & \textbf{5} & \textbf{M} & \textbf{SD} \\ \hline
Completeness & 0 & 0 & 1 & 14 & 9 & 4.33 & 0.56 \\ \hline
Consistency & 0 & 0 & 2 & 19 & 3 & 4.04 & 0.46 \\ \hline
Process Traceability & 1 & 2 & 1 & 10 & 10 & 4.08 & 1.10 \\ \hline
Feasibility & 0 & 1 & 10 & 10 & 3 & 3.63 & 0.77 \\ \hline
Representational Traceability & 0 & 0 & 2 & 9 & 13 & 4.46 & 0.66 \\ \hline
Conformance & 0 & 0 & 3 & 14 & 7 & 4.17 & 0.64 \\ \hline
Overall Effectiveness & 0 & 0 & 3 & 10 & 11 & 4.33 & 0.70 \\ \hline
\end{tabular}
\end{table}

\section{Participant Task Completion Times}

Table~\ref{tab:completion_time} shows the task completion times (in minutes) for each participant under both approaches. 
Each column corresponds to one participant (01–24). 
The bottom row reports the mean and standard deviation across all participants.

\newpage

\begin{table*}[h!]
\centering
\caption{Participant task completion times by approach (minutes)}
\label{tab:completion_time}
\begin{tabular}{|P{2.5cm}|*{24}{P{0.45cm}|}}
\hline
\textbf{Participant ID} & \textbf{01} & \textbf{02} & \textbf{03} & \textbf{04} & \textbf{05} & \textbf{06} & \textbf{07} & \textbf{08} & \textbf{09} & \textbf{10} & \textbf{11} & \textbf{12} & \textbf{13} & \textbf{14} & \textbf{15} & \textbf{16} & \textbf{17} & \textbf{18} & \textbf{19} & \textbf{20} & \textbf{21} & \textbf{22} & \textbf{23} & \textbf{24} \\ \hline
\textbf{Approach 1 Time} & 40 & 30 & 35 & 30 & 25 & 20 & 30 & 40 & 18 & 40 & 20 & 30 & 20 & 20 & 25 & 20 & 50 & 25 & 30 & 28 & 30 & 35 & 35 & 23 \\ \hline
\textbf{Approach 2 Time} & 25 & 56 & 40 & 35 & 30 & 25 & 30 & 30 & 26 & 20 & 40 & 30 & 40 & 30 & 30 & 40 & 35 & 25 & 17 & 28 & 30 & 25 & 20 & 27 \\ \hline
\textbf{Time Difference} & 15 & -26 & -5 & -5 & -5 & -5 & 0 & 10 & -8 & 20 & -20 & 0 & -20 & -10 & -5 & -20 & 15 & 0 & 13 & 0 & 0 & 10 & 15 & -4 \\ \hline\multicolumn{25}{r}{\textit{Overall mean (Approach 1 = 29.13 $\pm$ 8.15 min, Approach 2 = 30.58 $\pm$ 8.38 min, Difference = -1.46 $\pm$ 12.47 min)}} \\[-0.8em]
\end{tabular}
\end{table*}


\end{document}